\documentclass[12pt,preprint]{aastex}
\usepackage{graphicx}
\title{A Multi-Component Analysis Indicates a Positronic Major Flare in GRS 1915+105}
\author{Brian Punsly\altaffilmark{1}}
\altaffiltext{1}{1415 Granvia Altamira, Palos Verdes Estates, CA USA 90274
and ICRANet, Piazza della Repubblica 10 Pescara 65100, Italy,
brian.punsly@verizon.net or brian.punsly@comdev-usa.com}

\begin{document}
\begin{abstract} A modeling strategy that is adapted to the study of
synchrotron-self absorbed plasmoids that was developed for the
quasar, Mrk 231, in Reynolds et al (2009) is applied to the
microquasar GRS 1915+105. The major flare from December 1993 shows
spectral evidence of three such self-absorbed components. The
analysis yields an estimate of the power that is required to eject
the plasmoids from the central engine that is independent of other
estimates that exist in the literature for different flares. The
technique has an advantage since the absorbed spectrum contains an
independent constraint provided by the optical depth at each epoch
of observation. The modeling procedure presented here
self-consistently determines the dimensions of the radio emitting
plasma from the spectral shape. Thus, structural dimensions are
determined analytically that can be much smaller than interferometer
beam-widths. A synthesis of the time evolution of the components
allows one to address the fundamental uncertainties in previous
estimates. First, the plasma is not protonic, but it is comprised of
an electron-positron gas. The minimum electron energy is determined
to be less than six times the electron rest mass energy. The
analysis also indicates that the plasmoids are ejected from the
central engine magnetically dominated. The temporal behavior is one
of magnetic energy conversion to mechanical energy as the plasmoids
approach equipartition. The time dependent models bound the
impulsive energy flux, $Q$, required to eject the individual major
flare plasmoids from the central engine to, $4.1 \times
10^{37}\mathrm{erg/s}< Q < 6.1 \times 10^{38} \mathrm {ergs/s}$.
\end{abstract}

\section{Introduction} It is possible that Galactic black hole accretion systems can
serve as laboratories in which AGN (active galactic nuclei) behavior
can be observed on time scales that are highly compressed. The black
hole candidate, GRS 1915+105 is of particular interest because of a
possible analogy with radio loud AGN. It routinely ejects powerful
radio emitting plasma out to large distances from the black hole at
relativistic speeds \citep{mir94,rod99,fen99,mil05}. There is no
other Galactic black hole that has produced nearly as many strong
radio flares that have been observed to be superluminal. The
propensity for high radio flux densities and superluminal ejections
make this Galactic black hole ideal for studying relativistic jet
formation. The superluminal ejections (jet) has been resolved into
multiple components at late times with each component limited in
size by the interferometer beamwidth \citep{fen99,dha00}. It is also
noteworthy that many of these states of high radio activity are
compact optically thick jets in which the X-ray spectrum is hard
\citep{dha00,rib04}. The X-ray luminosity of GRS 1915+105 is also
one of the highest of any known Galactic black hole candidate
\citep{don04}. Unlike other black holes in which the compact jet
occurs in a "low hard" spectral state, the X-ray luminosity of the
GRS 1915+105 can be very high ($\approx 20 \% $ of the Eddington
limit) in the hard state when a compact jet is present
\citep{fuc03}. It is not clear if the the two remarkable properties
of a luminous accretion flow and extraordinary superluminal
ejections are related. The existence of both properties make it
tempting to speculate that GRS 1915+105 is a scaled down version of
a radio loud quasar.
\begin{table}
\caption{Flux Densities of GRS 1915+105}
{\footnotesize\begin{tabular}{cccccc} \tableline\rule{0mm}{3mm}
Date &  UT Time & Frequency & Flux Density & Radio & Error\\
  &  (days) & (GHz) & mJy & Telescope & \% \\
\tableline \rule{0mm}{3mm}
12/6/93 & 0.02 & 1.465 & 209 & VLA-D  & 5\\
12/6/93 & 0.59 & 3.277 & 1083 & Nancay & 5  \\
12/6/93 & 0.02 & 4.885 & 503 & VLA-D & 5 \\
12/6/93 & 0.03 & 8.415 & 398 & VLA-D & 5  \\
12/6/93 & 0.03 & 14.965 & 241 & VLA-D & 5  \\
12/6/93 & 0.03 & 22.485 & 239 & VLA-D & 5   \\
12/6/93 & 0.60 & 234.000 & 123 & IRAM & 10   \\
\tableline \rule{0mm}{3mm}
12/11/93 & 0.57 & 1.407 & 1319 & Nancay  & 5 \\
12/11/93 & 0.38 & 2.695 & 860 & Bonn & 10 \\
12/11/93 & 0.57 & 3.277 & 901 & Nancay &  5  \\
12/11/93 & 0.50 & 10.550 & 403 & Bonn & 10  \\
\tableline \rule{0mm}{3mm}
12/14/93 & 0.56 & 1.407 & 653 & Nancay & 5   \\
12/14/93 & 0.60 & 1.410 & 615 & Bonn & 10   \\
12/14/93 & 0.85 & 1.465 & 650 & VLA-D  & 5 \\
12/14/93 & 0.56 & 3.277 & 646 & Nancay & 5   \\
12/14/93 & 0.85 & 4.885 & 390 & VLA-D  & 5 \\
12/14/93 & 0.85 & 8.415 & 270 & VLA-D  & 5 \\
12/14/93 & 0.85 & 14.965 & 170 & VLA-D  & 5 \\
12/14/93 & 0.85 & 22.480 & 130 & VLA-D  & 5 \\
\end{tabular}}
\end{table}

\par In order to understand the temporal
relationship between the episodic energy ejection mechanism and the
accretion state, one must be able to establish the magnitude of the
mechanical energy flux in these ejecta. The major flare in 1994 is
well chronicled \citep{mir94}. Unfortunately, minimum energy
estimates of the kinetic luminosity, $Q$, vary by orders of
magnitude $10^{37}\mathrm{erg/s}< Q < 7 \times 10^{41} \mathrm
{ergs/s}$ \citep{gli99,rod95,lia95,fen99}. This paper uses
techniques developed in \citet{rey09} to constrain these estimates
and remove the layers of assumptions. Previous efforts to estimate
jet power have been plagued by various unknowns. Each one of these
unknowns can independently contribute an order of magnitude or more
of uncertainty to the estimates. The four major ones are:
\begin{enumerate}
\item Is the plasma protonic? It is often assumed that the plasmoid is either protonic (normal
electron - proton gas) or positronic (a pair plasma). Proton kinetic
energy will typically dominate in a relativistic outflow. Major
flares have optically thin radio spectra at late times.
The energy of the synchrotron emitting leptons is then dominated
by low energy particles and the particle number density becomes
high. If the radio spectrum extends to low frequency and neutrality
is maintained by protons, the energy flux in the protons can be more
than two orders of magnitude larger than the leptonic energy flux.
\item What is the minimum electron energy, $E_{min}$? Again, this is given by fiat in previous estimates. If
$E_{min}=1$ in units of the electron rest mass energy, $m_{e}c^{2}$, then the radio spectrum will extend to
frequencies below those observed and the total number of particles increases dramatically for optically thin
radio emitting plasmoids. The value of $E_{min}$ is another parameter in which orders of magnitude of energy
uncertainty is hidden.
\item There is uncertainty in the size of the region that produces the bulk of the radio emission.
The kinetic luminosity of the plasma, $Q$, that produces the radio
emission scales with the size of the emitting region. The size of
the physical region that produces the preponderance of the radio
emission in major flares appears to be smaller than the FWHM (full
width at half maximum) beamwidth of the VLA, VLBA or MERLIN
interferometers as evidenced by the VLBA and MERLIN images in
\citet{rod99,dha00,fen99}. The existence of unresolved components,
significantly smaller than the beamwidth FWHM, would indicate that
the size of the emitting region has likely been typically
over-estimated in the past. Since the interferometer beamwidth is
potentially much larger than the plasmoid size, extremely liberal
sizes have been chosen a priori for energy estimates. Examples that
are mentioned in the Discussion section are:
\begin{itemize}
\item the entire deconvolved angular dimensions of an unresolved "jet" that was partially resolved from an adjacent radio component has been used as the size of the radio emitting plasmoid projected onto the sky plane in \citet{gli99,rod99}
\item and an assumed speed of light expansion of the plasmoid from an estimated initiation time was used as the plasmoid size in \citet{fen99}.
\end{itemize}
These could be considered as the most extreme upper bounds on the
plasmoid size that are compatible with a subset of the data rather
than estimates for the plasmoid size. For a minimum energy estimate
of a spherical emission region of radius, $r$, $Q \sim r^{9/7}$
\citep{lia95,fen99}. Thus, order of magnitude errors in the size of
the region of the preponderance of emissivity result in even larger
errors in the estimates for $Q$.
\item Is the minimum energy or the equipartition assumption justified? Such an assumption is typically made just
to obtain an extra constraint that can be used to restrict the
variables in the calculation of the energy flux. Yet, there is no
evidence that this assumption is justified.
\end{enumerate}

\par In \citet{rey09}, a powerful modeling method for analyzing relativistic plasma ejections was found in the context
of the quasar, Mrk 231. The modeling method exploits the fact that
synchrotron self-absorbed plasmoids are restricted physically by the
shape of their spectrum. In particular, the frequency and the width
of the spectral peak. This provides two added pieces of information
beyond the spectral index and flux density of the optically thin
high frequency tail (i.e., the only information that is used in
other estimates of the jet power of GRS 1915+105). Such an analysis
provides strong constraints on the size of the emitting region. In
order to implement the modeling method, one must have sufficient frequency
coverage so that the peak and the high frequency tail of the
spectrum are defined. Furthermore, the data must be
quasi-simultaneous because strong flares evolve rapidly, especially
in their compact self-absorbed phase. A literature search revealed
one such instance of broadband simultaneous frequency coverage in
\citet{rod95} for a flare in December 1993. The modeling of the
self-absorbed components of this flare and their time evolution is
the subject of this study. The next section describes the
synchrotron self-absorbed fits to the spectra. Section 3 provides
the range of physical parameters that are consistent with these
fits. Section 4 discusses the possibility of protonic plasma in the
synchrotron emitting region. Section 5 uses the time evolution
information to bound the power required to create the relativistic
ejecta.

\begin{figure}
\begin{center}
\includegraphics[width=95 mm, angle= 0]{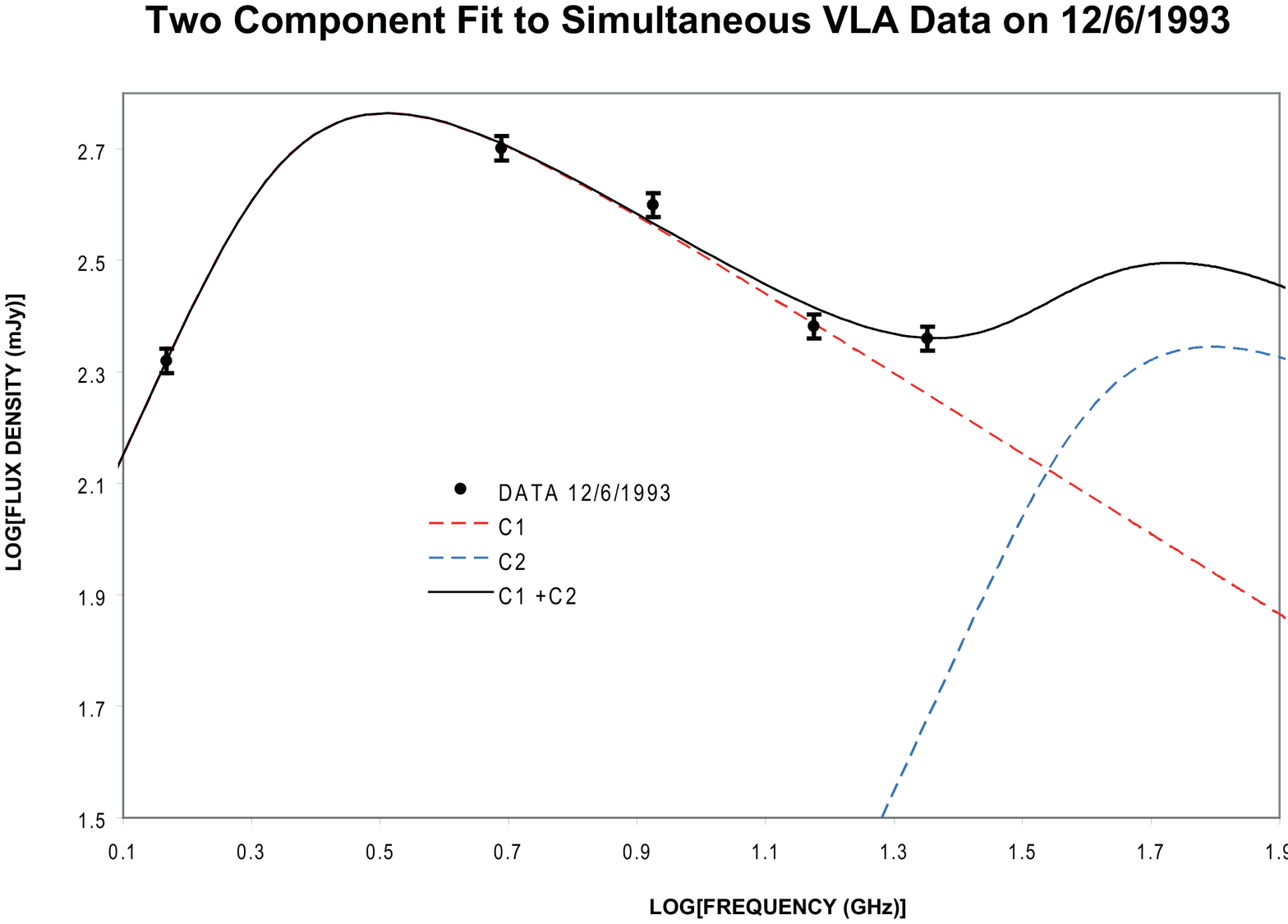}
\includegraphics[width=95 mm, angle= 0]{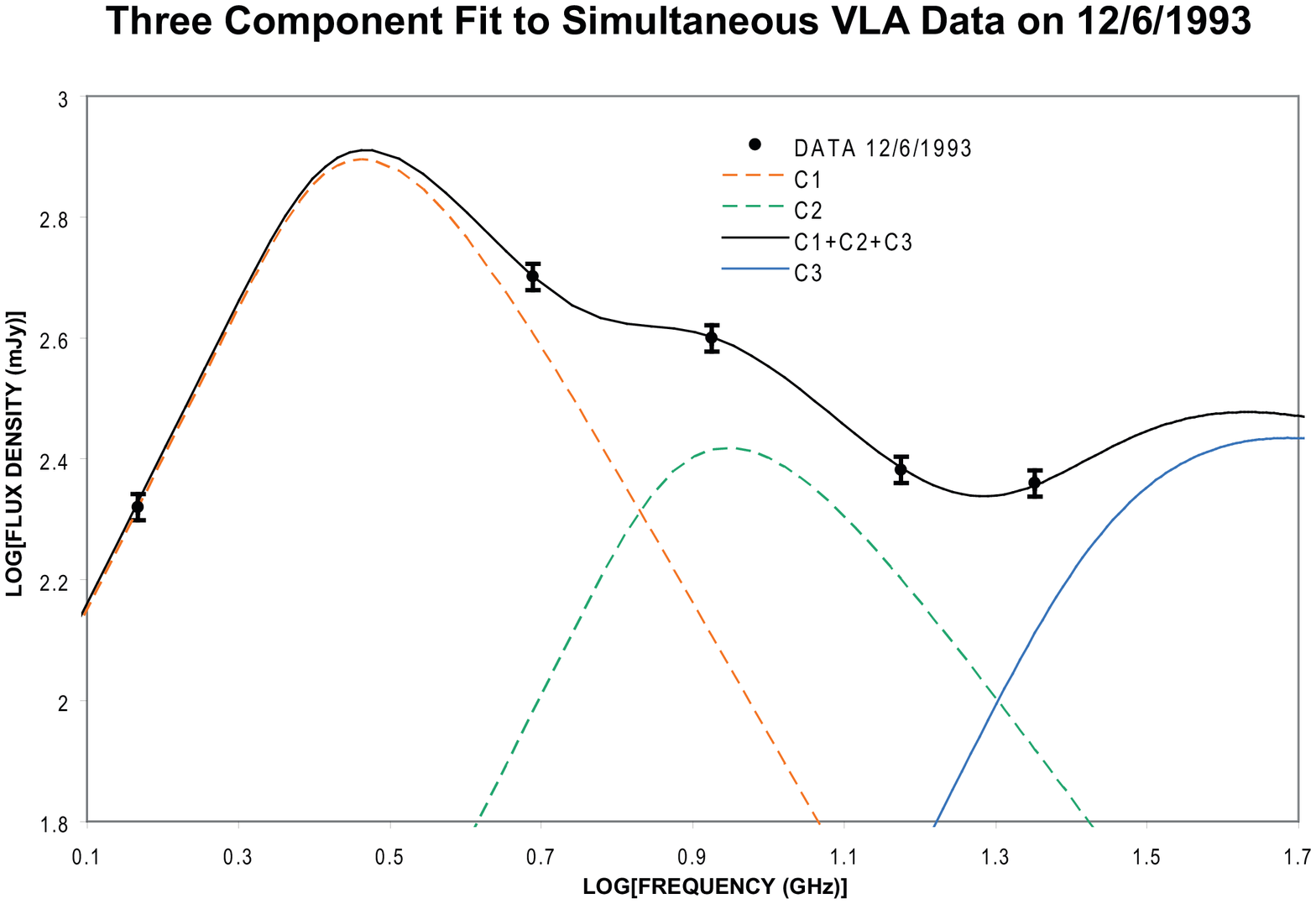}
\caption{A comparison of a two component fit and a three component
fit of the simultaneous VLA data on December 6.}
\end{center}
\end{figure}
\begin{figure}
\begin{center}
\includegraphics[width=125 mm, angle= 0]{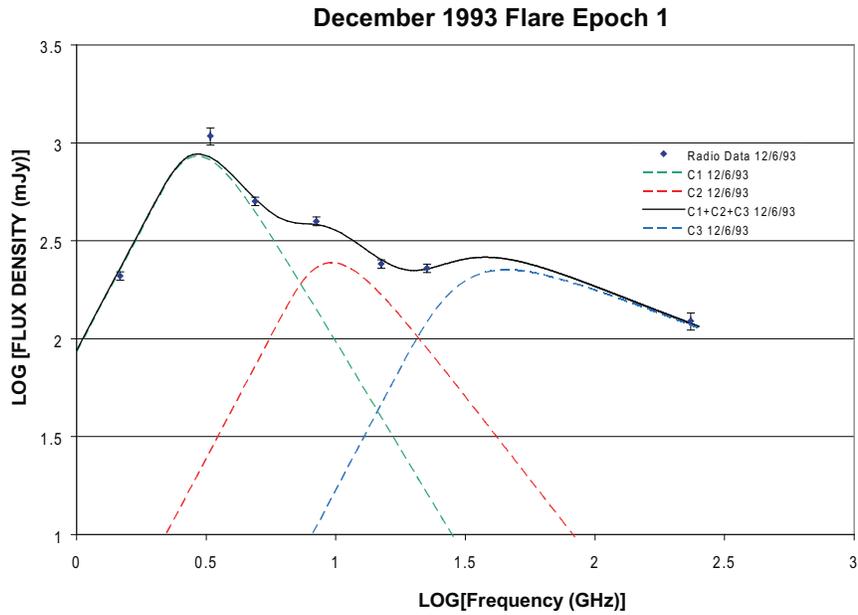}
\caption{The broad band quasi-simultaneous spectrum of GRS 1915+105
on December 6, 1993. The data is fit to three synchrotron
self-absorbed components. They are labeled C1, C2 and C3. The
component C1 is the lowest frequency peaked component and C3 the
highest. The 3.277 GHz data and the 234 GHz data is taken 12 hours
later than the other data on the plot.}
\end{center}
\end{figure}

\begin{figure}
\begin{center}
\includegraphics[width=125 mm, angle= 0]{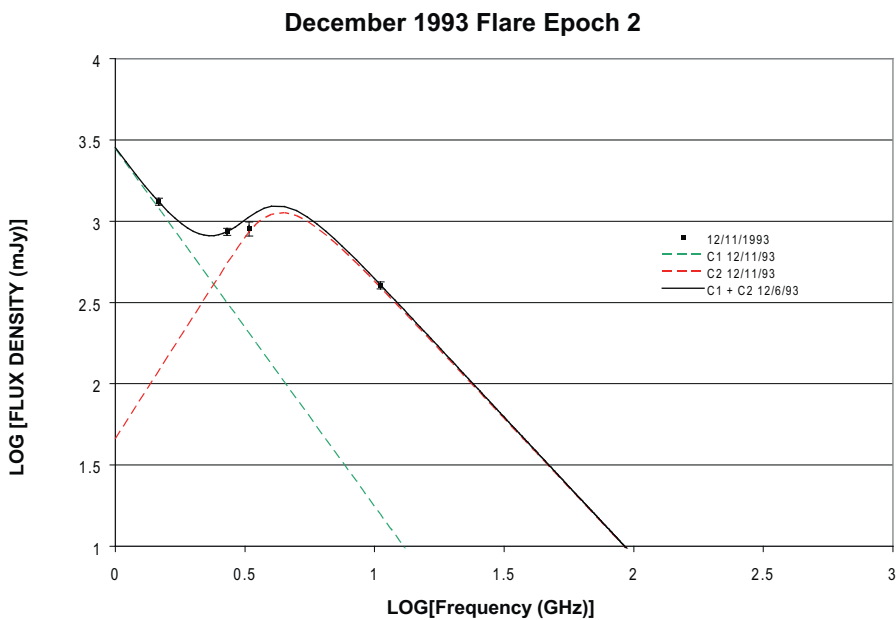}
\caption{The spectrum of GRS 1915+105 on December 11, 1993. The
frequency coverage is certainly more sparse than on December 6. But,
there is enough data to constrain a local maximum near 5 GHz that is
attributed to component C2. Since the frequency coverage ends at
10.5 GHz, there is an insignificant contribution expected from the
high frequency component, C3, to this data. Both C1 and C2 are
assumed to have the same value of $\alpha$ as for epoch December 6,
1993}
\end{center}
\end{figure}
\begin{figure}
\begin{center}
\includegraphics[width=125 mm, angle= 0]{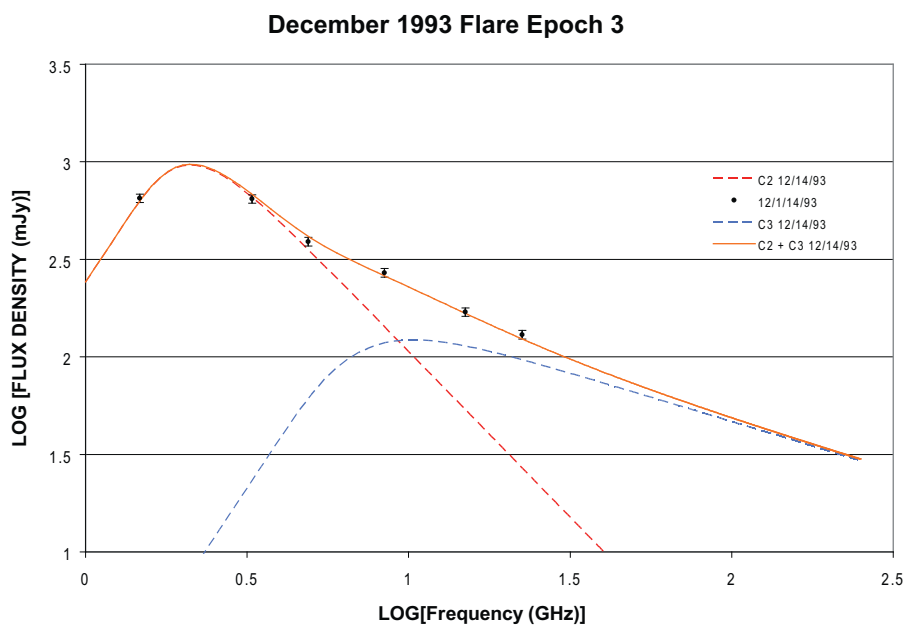}
\caption{The spectrum of GRS 1915+105 on December 14, 1993.
Component C1 has evolved to lower frequency and does not appear to
contribute noticeably near the local maximum near 2.5 GHz. The 6
data points fix the 6 free parameters in the two synchrotron-self
absorbed models of C2 and C3. The derived value of $\alpha$ is the
same in all three epochs for C2.}
\end{center}
\end{figure}
\begin{figure}
\begin{center}
\includegraphics[width=125 mm, angle= 0]{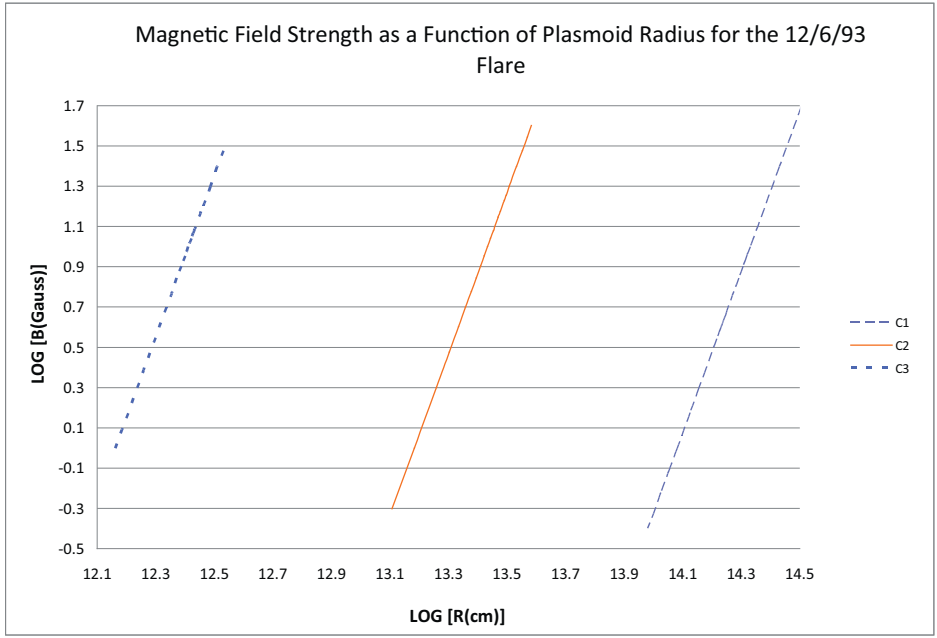}
\caption{The magnetic field strength, $B$, as a function of the
radius of the model for each of the three components on December 6.
Notice that $B$ increases with increasing $R$.}
\end{center}
\end{figure}

\begin{figure}
\begin{center}
\includegraphics[width=125 mm, angle= 0]{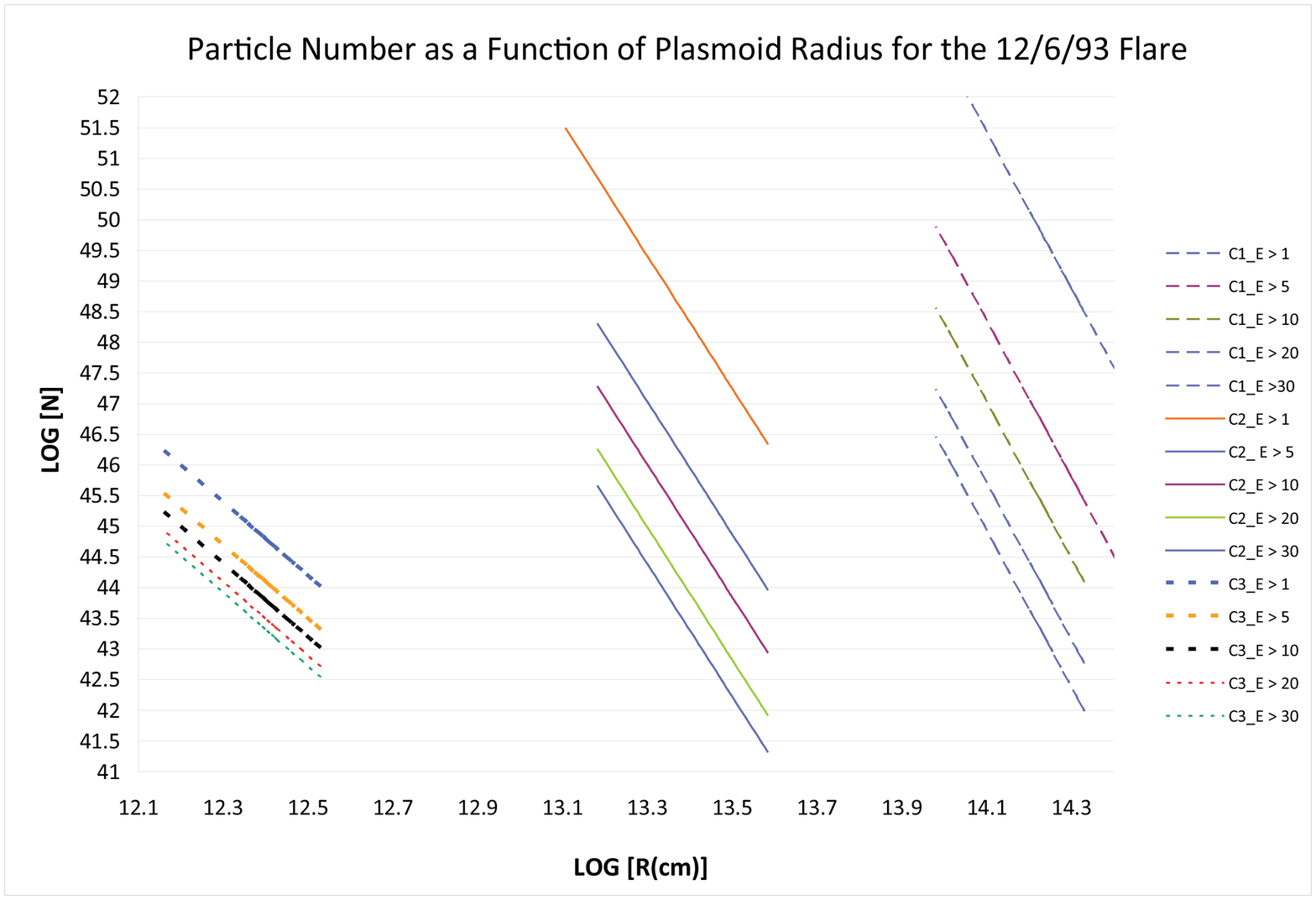}
\caption{The total number of particles in the plasmoid, $N$, as a
function of the radius of the model for each of the three components
on December 6. Notice that $N$ decreases with increasing $R$. The
value of $N$ depends on $E_{min}$. If the optically thin region of
the spectrum is very steep then the dependence on $E_{min}$ is
dramatic. This paper explores 5 models with various choices of
$E_{min}$ set at the values 1, 5,10, 20 and 30.}
\end{center}
\end{figure}
\section{The Major Flare of December 1993} The radio monitoring of \citet{rod95} revealed
that a large flare had emerged from GRS 1915+105 on December 6,
1993. The flare was of comparable in magnitude to the large flare of
March 1994 described in \citet{mir94}. The December 1993 flare was
unique in that there was spectral coverage all the way from 1.465
GHz to 234 GHz. The data from \citet{rod95} is repeated in Table 1.
There are three dates that are well sampled.There are 7 data points
for December 6, 4 data points for December 11, and 8 data
points for December 14 for a total of 19 data points. This a bit
deceiving because 3 data points on December 14 are all at $\approx$
1.4 GHz. Thus there are really only 6 independent data points on
December 14, for a total of 17 independent data points on the three
days. Notice that all 3 of the 1.4 GHz observations on December 14
are the same within the quoted errors. This indicates that the
optically thin spectrum was relatively stable over time scales of at
least 4.5 hours.
\par In this section, fits to the data in Table 1 are presented using synchrotron self absorbed
components. Figures 1 and 2 show the flux density data and some
spectral fits on December 6, 1993. Inverted spectra in self-absorbed
radio sources are common. However, the inverted spectrum from 1.465
GHz to 3.277 GHz juxtaposed to the steep spectrum, $\alpha =1.9$
(where the convention $L_{\nu}=L_{0}\nu^{-\alpha}$ is assumed), from
3.277 GHz to 4.885 GHz represents very large gradients in $\alpha$
with frequency. Such a large gradient (equivalently, such a narrow
local maximum in the spectrum) is extremely unusual for a radio
source and provides a strong constraint on the nature of the flare
that will be exploited in this section.
\subsection{Synchrotron Self Absorbed Component Models} The peaked spectral luminosity of ejected plasmoids is generally
considered to be a consequence of radiative transfer through a synchrotron self absorbed (SSA) plasma.
It is assumed that the emission region is a homogeneous sphere
ejected from the central engine. This is certainly more accurate
than a continuous jet for emission that shows no evidence in
\citet{rod95,rod99,rus10,fos96,fen99,rus11} of being quasi-steady.
The lack of protracted periods of powerful quasi-steady flux is in
contradiction to what would be expected from a large smoothly
varying structure, like the jet models in \citet{kon81}. Every
indication is that the preponderance of the emission on December 6,
1993 is in the form of a large impulsive burst possibly on the
background of some low luminosity feature (perhaps the continuous
jet). The episodic flaring nature of the source indicated in
\citet{rod95,poo97,rus10,rus11} is consistent with large ejections
of highly energized plasma, with the epoch December 6, 1993 being
one of the most energetic. The epoch December 6, 1993 plasmoid might
have an irregular shape with an inhomogeneous density, but without
any prior knowledge of these geometric properties, a spherical solid
is implemented as a zeroth order approximation. The strategy is to
perform the final calculation in the plasma rest frame using known
variables from observation. Observed quantities will be designated
with a subscript, ``o", in the following expressions. Taking the
standard result for the SSA attenuation coefficient in the plasma
rest frame and noting that $\nu = \nu_{o} / \delta$, we find from
\citet{rey96,gin69},
\begin{eqnarray}
&& \mu(\nu)=\frac{3^{\alpha +
1}\pi^{0.5}g(p)e^{2}N_{\Gamma}}{8m_{e}c}\left(\frac{eB}{m_{e}c}\right)^{(1.5
+
\alpha)}\nu_{o}^{(-2.5 + \alpha)} \delta^{(2.5 + \alpha)}\;,\\
&& g(n)= \frac{\Gamma[(3n + 22)/12]\Gamma[(3n + 2)/12]\Gamma[(n +
6)/4]}{\Gamma[(n + 8)/4]}\;.
\end{eqnarray}
The Doppler factor, $\delta$, is given in terms of $\Gamma$, the
Lorentz factor of the outflow; $\beta$, the three velocity of the
outflow and the angle of propagation to the line of sight, $\theta$;
$\delta=1/[\Gamma(1-\beta\cos{\theta})]$ \citep{lin85}. This
equation derives from an assumed powerlaw energy distribution for
the relativistic electrons, $ N(E)= N_{\Gamma}E^{-n}$, where the
radio spectral index $\alpha = (n-1)/2$ and $E$ is the energy of the
electrons in units of $m_{e}c^{2}$. The radiative transfer equation
was solved in \citet{gin69} to yield the following
parametric form for the observed flux density, $S_{\nu}$, from the
SSA source,
\begin{eqnarray}
&& S_{\nu} = \frac{S_{o}\nu^{-\alpha}}{R\mu(\nu)} \times \left(1 -
e^{-\mu(\nu) R}\right)\;,
\end{eqnarray}
where R is the radius of the spherical region in the rest frame of
the plasma and $S_{o}$ is a normalization factor. In the spherical,
homogeneous approximation, one can make a simple parameterization of
the SSA attenuation coefficient, $\mu(SSA)=\mu_{o}\nu_{o}^{(-2.5 +
\alpha)}$. If one assumes that the source is spherical and
homogeneous then there are three unknowns in equation (3),
$R\mu_{o}$, $\alpha$ and $S_{o}$, and $S_{\nu}$ is determined by
observation.

\par Figure 1 shows the simultaneous VLA data from December 6, 1993. The
data seems to indicate three local maxima in the spectrum (one
between 1.4 GHz and 5 GHz, one near 8.3 GHz and one near 22 GHz). It
should be noted that the SSA function is not an arbitrary shape. For
example, it has precisely one maximum. Thus, even though each SSA
function has three parameters, there is no guarantee, in general,
that two SSA components can fit six or less points exactly. Two SSA
components can create at most two local maxima in spectrum. Thus a
minimum of three components seems indicated by the simultaneous
subset of the data. The comparison between a two component fit and a
three component fit is given in Figure 1. The fits are made by
letting all the SAA component parameters vary, then minimizing the
sum of the squares of the residuals between the simultaneous data
and the models. The deviation of the two component model and the
data is beyond what is expected based on the statistical uncertainty
in the data. Each VLA data point can be treated as an independent
statistical variable, $X_{i}$, $i=1,2,3,4,5$ that is normally
distributed with a mean and standard deviation given by the flux
density value and the flux density error in Table 1, respectively.
The two pints that are more than 1$\sigma$ from the mean only have
about a 10\% chance of occurring by random chance and indicate the
inability to fit the local maximum in that region. Consider the null
hypothesis that the five data points are fit by the model. Each data
point provides an independent statistical variable and probability.
Therefore, comparing the models in Figure 1 to the data implies that
the probability of rejecting the null hypothesis is 0.99 for a two
component fit and 0.512 for a three component fit. Thus, the three
component fit cannot be rejected on a statistical basis, but the two
component fit is rejected with a very high statistical significance.
Restricting the discussion to the simultaneous VLA data removes two
uncertainties for the fits of the December 6 data in Table 1. First
of all, the data is simultaneous, so there is no issue of
variability within the short observations. Secondly, the VLA data is
very high signal to noise so the errors are small (see Table 1)
which allows us to exclude a two component model.
\par In Figure 2, the entire data-set from December 6 is fit by using the three
parameter SSA spectrum described above in equation (3) to represent
each spectral inflection point. There appears to be three compact
components corresponding to the three local maxima. The component
that is constrained the least is C3, the high frequency component,
since it is highly dependent on the 234 GHz flux density that was
taken 12 hours later than the VLA data. The radio flux in GRS
1915+105 can vary significantly in 12 hours \citep{rod95,rus10}. The
other less reliable point is the 3.277 GHz datum from Nancay which
was also taken 12 hours after the VLA data. Thus, its exact value
was not determinant in fitting component C1. It appears that the
flux density was likely increasing near the spectral peak during
those 12 hours (the flare was still rising). The fit in Figure 2
uses 9 free parameters to fit 7 data points. The fit that was chosen
is not unique. For example, one could pick the fit that minimizes
the sum of squares of the errors. Instead, based on the above
comments on the 12 hour delay in collecting the 234 GHz and 3.277
GHz data, these data were weighted less heavily in selecting the
fit.

\par Before, discussing the other epochs, it is important to digress on the issue of
solving for all three epochs simultaneously and its relationship to
the number of total unknowns and the total number of free parameters
in the models. The fact that all the free parameters are constrained
for C2 and C3 in Figure 4 (epoch December 14, 1993), can be used to
remove the arbitrariness of the fits for December 6 and 11 in Table
1. In particular, these values of $\alpha$ for C2 and C3 can be used
in the December 6 and 11 fits to the data. Similarly, the steep low frequency spectral index seen in
the December 11 data in Figure 3 can be used to determine $\alpha$
in C1 in both that epoch and on December 6. In summary, using the
fact that there is a causal connection between the plasma state in
the 3 epochs is being utilized to reduce the total number of free
parameters in the 3 epochs from 21 to 17. There are 17 data points
in the three epochs and 17 unknowns which is a well defined
solution.

\par The flare is monitored for a couple of weeks in \citet{rod95} with varying spectral coverage.
Table 2 includes the parameters for the fits to the data from two
other epochs of observation of the same flare. On December 11, 1993
and December 14, 1993 there was sufficient spectral coverage to
resolve component C2. Figures 3 and 4 show the data from those
epochs and the best fits. Clearly there is spectral evolution. The
data from December 11, 1993 does not have sufficient high frequency
coverage to fit the contribution of component C3. Thus, there is no
reference to this component on December 11 in the table. The
spectral turnover of component C1 has evolved on December, 11, due
to component expansion, so as to fall below the observed frequency
range at 1.4 GHz as indicated in the table. On December 14, there is
insufficient low frequency coverage to put any constraints on
component C1 that had expanded so as to be optically thin far below
1.4 GHz and was apparently very weak at the lowest observed
frequencies. Figure 3 has four data points and is fit with six
parameters arising from the parametric form (defined in equation 3)
for the SSA spectra of the components C1 and C2. This is reduced to
four free parameters by assuming that $\alpha$ is unchanged from
December 6 (see the discussion above). In Figure 4, there are 6 data
points that are fit by the 6 free parameters associated with SSA
spectral fits to the components C2 and C3. The fits have the same
values of $\alpha$ as they had on December 6.
\par Figures 1 to 4 show the time evolution of the three components based on the fits at the three epochs.
Only component, C2, is adequately sampled so as to be resolved in
all three epochs. Figures  1 to 4 do not tell us very much about the
evolution of component C1 since the spectral peak has evolved below
the data sampling range by the second epoch. It is clear from
Figures 1 to 4 that component C2 is not adiabatically expanding. If
that were true the spectral shape would merely migrate toward lower
frequency with some decrease in the peak flux \citep{mof75}. To the
contrary, the peak flux density increased dramatically from December
6, 1993 to December 11, 1993. This indicates that the plasmoid is
evolving to a state of higher radiative efficiency. The evolution
between December 11 and December 14 is closer to adiabatic expansion
as the peak flux is approximately constant. Thus, the rate at which
the radiative efficiency is increasing slowed down dramatically
after December 11. The highest radiative efficiency is in the
minimum energy configuration. So, the plasmoid was far from the
minimum energy configuration on December 6, but appears to be
evolving towards minimum energy after December 11. By contrast,
Figure 2 and 4 indicate that component C3 is shifting in frequency
and the spectral peak is weakening after December 6. This type of
behavior is similar to that of a component evolving very close to
adiabatic expansion \citep{mof75}. Note that C3 is a much weaker
component than C1 and C2.
\subsection{Spectral Ageing and the Constant $\alpha$ Assumption}
The constant $\alpha$ assumption is required to formally close the
set of equations as discussed above. In this subsection, the implications
of this assumption in terms of spectral ageing and the generality of the model is
discussed.
\par Note that the mathematical
constraint of "constant $\alpha$" translates physically to the
notion that after December 6, the spectral ageing of the components
is not dramatic. This does not mean that there was not pronounced
spectral ageing before December 6 and modest spectral ageing after
December 6. There really is no direct information on the time
history of the ejecta before December 6. For example, the different
spectral indices of the three components on December 6 is consistent
with the components being ejected staggered in time: C1 first and C3
last with the spectral steeping occurring from radiation losses. If
the spectral ageing mechanism is similar for the three components
then in any time snapshot C1 will be steeper than C2 and C2 steeper
than C3. A simple reverse time evolution of the models found in
section 5, indicate substantial spectral ageing must have occurred
due to synchrotron losses in a strong internal magnetic field and
inverse Compton losses in a strong ambient X-ray field. This is a
very plausible explanation of the unusually steep spectral indices
of the components C1 and C2 from December 6 to December 14. For
example, consider the existence of a strong X-ray flare during the
time that component C1 was launched that decayed on a time scale of
hours. Inverse Compton losses in the first hour after ejection would
make the leptons in C1 cool more than the leptons in C2 and those in
C3 might not be cooled very much at all (because the X-ray flare was
weak when C3 was launched). So in principle, for a time variable
source, not only is the elapsed time for spectral ageing important,
but also the strength of the ambient X-ray background at the time at
which the plasma was ejected. Similarly, one might have a scenario
in which a magnetic flare in the inner accretion flow launches the
jet. The first ejection removes a large fraction of the magnetic
flux. Most of the remainder is removed by the ejection of C2 and C3
has a relatively weak magnetic field. This scenario would also
create different spectral ageing rates in the spectral components.
The leptons would cool primarily by synchrotron self-Compton
emission in the early stages and synchrotron emission later on. The
most rapid rate of ageing would also be in C1. As the plasmoids
expand, the rate of spectral ageing would decrease since the
magnetic field is getting weaker.
\par The constant $\alpha$ assumption noted above is tantamount
to the following scenario. The plasmoids are ejected and cooled in
different ambient X-ray backgrounds, or magnetic environments. In
particular, the X-ray background or magnetic field was strong when
C1 was launched and decayed during the time interval that subsequent
ejecta were emitted. Thus, one day after being ejected, the leptonic
energy spectrum of C1 is steeper than the leptonic energy spectrum
of C2 would be one day after it was ejected. By December 6, the
spectral ageing rate is comparatively very slow (a much weaker
source of cooling and most of the high energy particles have already
been cooled). This "conjecture" is consistent with the models found
in section 5. After December 6, the models are shown in section 5 to
have higher synchrotron cooling rates than inverse Compton cooling
rates. Thus, after December 6, one would expect that synchrotron
cooling is the dominant driver of spectral ageing. The total energy
emitted in the synchrotron spectrum between December 6 and December
14 is $\approx 4 \times 10^{39}$ ergs and the total energy stored in
the leptons comprising the ejected plasmoids in the models of
section 5 is about $\approx 6 \times 10^{42}$ ergs. This indicates
that the radiation losses are a negligible contributor to the energy
budget of the plasmoids after December 6. Thus, spectral ageing
after December 6 is not be expected to be large, self-consistently
with the models.
\par The data on December 11 and 14 is not sufficiently sampled to show that further spectral ageing has or has not occurred
after December 6. If indeed there is some spectral ageing, this
should not drastically affect the results of the models. The
empirical reason is that the basic extreme inflections in the
spectral shape drive the model. Namely, the approximate values of
$\alpha$ are still determined directly by the data. The data clearly
indicates a steep energy index for C1 and C2 per the second
paragraph of this section. Also, the energy spectrum of C3 cannot be
steep due to the 234 GHz point on December 6 and the 22.5 GHz point
on December 14. It must be remembered that because of the constant
spectral index expedience that the solutions presented here are not
unique. It is however argued that significant departures from the
spectral indices used in the analysis are not supported by the data
as discussed above. The arguments presented are not a rigorous proof
that there is no significant spectral ageing after December 6, but
there is no direct evidence to say otherwise.

\section{Kinematic Models of the Plasmoids}
In this section, the fit parameters in Table 2 are related to actual
physical parameters. There is a finite range of physical parameters
that are consistent with these spectral fits. To make the
connection, one needs to relate the observed flux density in
equation (3) to the local synchrotron emissivity within the plasma.
The synchrotron emissivity is given in \citet{tuc75} as
\begin{eqnarray}
&& j_{\nu} = 1.7 \times 10^{-21} (4 \pi N_{\Gamma})a(n)B^{(1
+\alpha)}(4
\times 10^{6}/ \nu)^{\alpha}\;,\\
&& a(n)=\frac{\left(2^{\frac{n-1}{2}}\sqrt{3}\right)
\Gamma\left(\frac{3n-1}{12}\right)\Gamma\left(\frac{3n+19}{12}\right)
\Gamma\left(\frac{n+5}{4}\right)}
       {8\sqrt\pi(n+1)\Gamma\left(\frac{n+7}{4}\right)} \;.
\end{eqnarray}

\par
One can transform this to the observed flux density, $S(\nu_{o})$,
in the optically thin region of the spectrum using the relativistic
transformation relations from \citet{lin85},
\begin{eqnarray}
 && S(\nu_{o}) = \frac{\delta^{(3 + \alpha)}}{4\pi D_{L}^{2}}\int{j_{\nu}^{'} d V{'}}\;,
\end{eqnarray}
where $D_{L}$ is the luminosity distance and $j_{\nu}^{'}$ is
evaluated in the plasma rest frame at the observed frequency. The
distance to GRS 1915+105 is debatable with estimates in the range of
6 kpc to 12.5 kpc \citep{mil05}. This estimate directly affects the
inferred rates that kinematic components are expanding and therefore
the determination of $\delta$. We adopt the popular values that
\citet{rod95,rod99} found to be consistent with the observations of
the flare from March 1994,
\begin{eqnarray}
&& D_{L} = 12.5 \,\mathrm{kpc}\;,\delta = 0.57 \;.
\end{eqnarray}
\begin{figure}
\begin{center}
\includegraphics[width=125 mm, angle= 0]{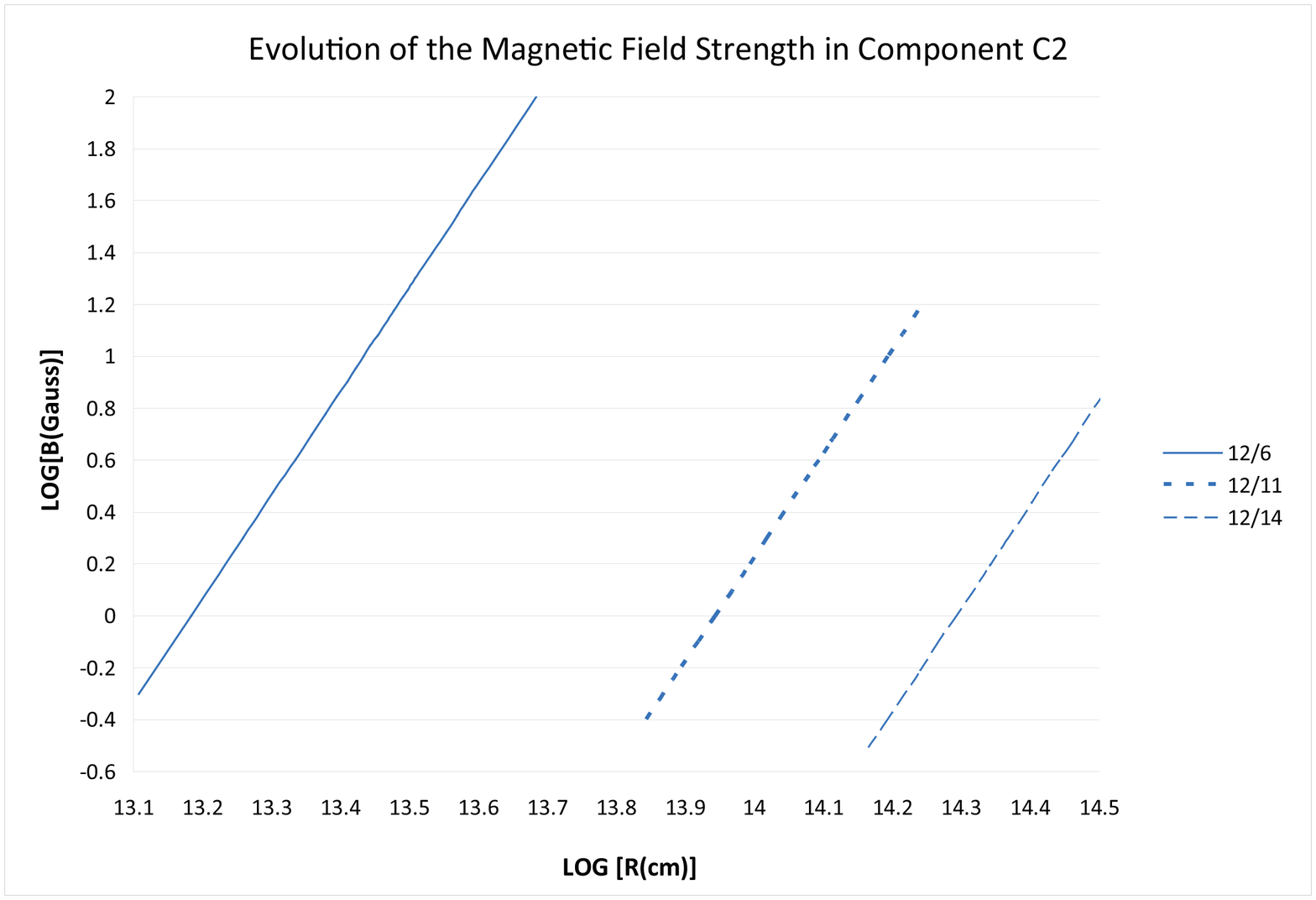}
\caption{The magnetic field strength, $B$, as a function of the
radius of the model for component C2 at three epochs, December 6,
December 11 and December 14. Notice that $B$ increases with
increasing $R$ at all epochs.}
\end{center}
\end{figure}

\begin{figure}
\begin{center}
\includegraphics[width=125 mm, angle= 0]{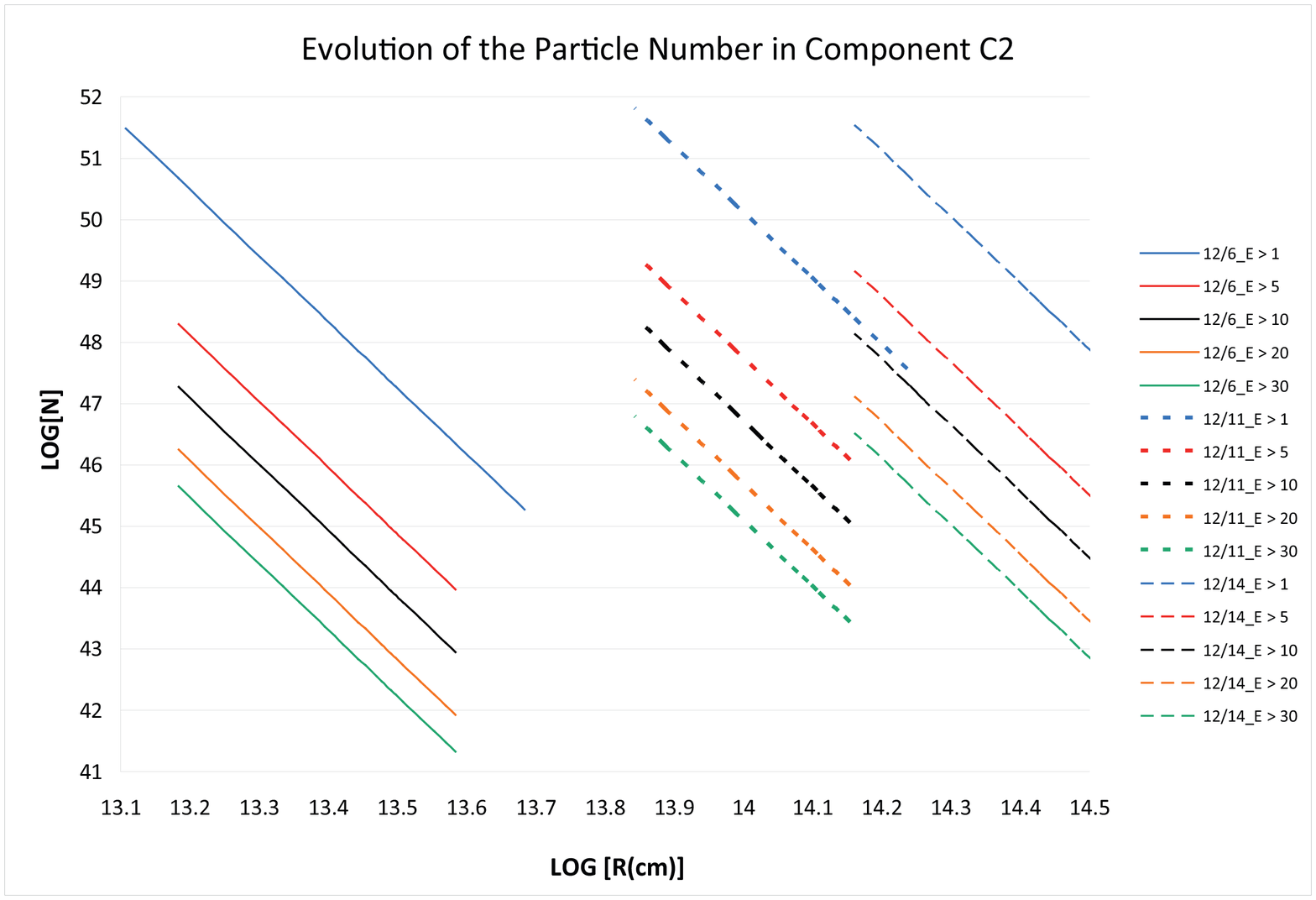}
\caption{The total number of particles in the plasmoid, $N$, as a
function of the radius of the model for for component C2 at three
epochs, December 6, December 11 and December 14. Notice that $N$
decreases with increasing $R$ at all epochs. This paper models 5
choices of the parameter $E_{min}$: 1, 5,10, 20 and 30.}
\end{center}
\end{figure}
\par Combining the results of equations (1) through (7) with the fits in Table 2, we have
two equations (constraints) and four unknowns ($\alpha$ is known
through Table 2). The constraints are given by frequency of the
spectral peak and peak flux density (equivalently, the normalization
of the background synchrotron power law and the frequency of unit
optical depth defined by $\mu({\nu})R=1$). The unknowns can be
described by $B$, the magnetic field, $N_{\Gamma}$, the
normalization of the energy distribution of leptons, $E_{min}$, the
minimum energy of the lepton energy power law and $R$, the radius of
the spherical plasmoid. The solution space is illustrated
graphically. Figures 5 and 6 are plots of $B$ and $N$, the total
number of particles in the plasmoid, as a function of R for each
value of $E_{min}$ for the three components on December 6, 1993,
where

\begin{eqnarray}
&& N= \int_{E_{min}}^{E_{max}}{N_{\Gamma}E^{-n}\, dE} \;.
\end{eqnarray}
Five representative values of $E_{min}$ are considered in the family
of models presented in the following that are equal to 1, 5, 10, 20
and 30. The integral in equation (8) is insensitive to the value of
$E_{max}$ for large $n$ as long as $E_{max}\gg E_{min}$. A liberal
value of $E_{max} = 10^{6}$ was chosen. Reasonable variations in
this choice will make modest changes to the energy in C3, since
$n=2$ is not large. However, the energy in this component appears to
be very reduced compared to the other components, regardless of this
choice. The models of C3 are the most poorly constrained because of
the sparse high frequency coverage. The trends to note are that $B$
always increases dramatically as the radius of the model of the
plasmoid increases for all three components. Furthermore, $B$, is
not a function of $E_{min}$. By contrast, $N$ decreases dramatically
as the radius of the plasmoid increases. The total number of
particles is very dependent on $E_{min}$ especially for C1 and C2
that have the steep high frequency power laws (equivalently, a steep
lepton energy distribution). It is interesting to note the size of
the largest component, C1. The highest resolution ever used to view
an ejected plasmoid in a major flare are the 2.0 cm VLBA observation
of \citet{dha00}. It is not clear from \citet{dha00} if the 2 cm
observations reveal the emitted component if the data is not
tapered. The VLBA observations at 2.0 cm are capable of resolving
components $R > 2 \times 10^{14}$ cm. Thus, only the largest models
of C1 could be marginally resolved by the VLBA and C2 and C3 could
not be resolved on December 6. The \citet{dha00} observations of two
different flares were unable to resolve the ejected region of high
emissivity in the image plane either with full resolution at 3.6cm
or with tapering at 2 cm, consistent with the predications of the
models presented here. Thus, the plasmoids that are considered here
are too small to have their sizes accurately fit with state of the
art interferometry, except possibly if they are observed at very
high frequency with VLBA.
\begin{figure}
\begin{center}
\includegraphics[width=125 mm, angle= 0]{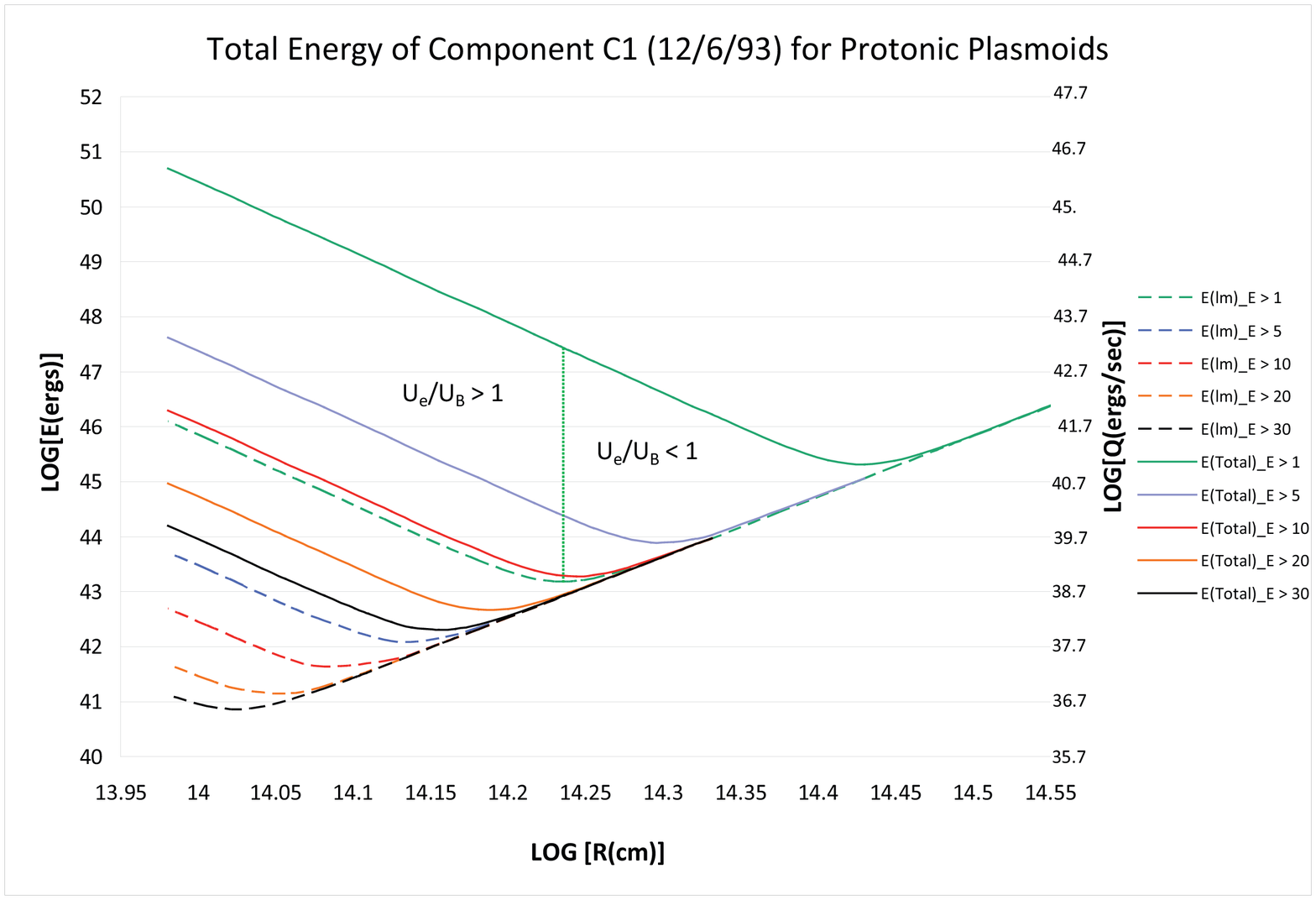}
\caption{The left hand axis is the scale for the total energy
stored, $E(\mathrm{Total})$, in C1 on December 6 for various
protonic models of the plasmoid as a function of plasmoid radius,
$R$. These values are contrasted with the lepto-magnetic energy, the
energy stored in the leptons and the magnetic field,
$E(\mathrm{lm})$. The protonic energy is dominant except for large
radii in which the magnetic field is large and particle number is
small, see Figures 5 and 6. The vertical dotted green line connects
the minimum energy lepto-magnetic solution for $E_{min}=1$ (the
dashed green curve) to the total energy of the protonic solution
with $E_{min}=1$ (the solid green curve). The value on the solid
curve is much larger indicating that most of the energy is in the
form of protonic kinetic energy for this "minimum energy solution."
The right hand axis is the scale for the power required to energize
and eject the plasmoid, C1, from the central engine as derived from
the models of December 6. The total power, $Q(\mathrm{Total})$, is
plotted assuming a protonic plasma. The power required to energize
and eject just the lepto-magnetic component, $Q(\mathrm{lm})$, is
also plotted separately for comparison. The horizontal lines can be
used to find the $Q(\mathrm{Total})$ that is associated with the
minimum energy solutions (minimum $Q(\mathrm{lm})$) for each value
of $E_{min}$.}
\end{center}
\end{figure}
\begin{table}
\caption{Description of Synchrotron-Self Absorbed Components}
{\footnotesize\begin{tabular}{ccccccc} \tableline\rule{0mm}{3mm}
Component &  Date & $\alpha$ & $S_{o}(\mathrm{ergs/cm^{2}-sec})$ & $R\mu_{o}$& Peak (GHz)& Peak (mJy)\\
\tableline \rule{0mm}{3mm}
C1 & 12/6/93 & 2.2 & $9.65 \times 10^{-3} $ & $5.30 \times 10^{-4}$  & 2.9& 857 \\
C1 & 12/11/93 &2.2 & $1.82 \times 10^{-3}$ & ... & $<1.4 $ & ...  \\
C2 & 12/6/93 & 1.7 & $2.60 \times 10^{-7}$& $1.50 \times 10^{-1}$ & 9.5 & 245  \\
C2 & 12/11/93 & 1.7 & $3.12 \times 10^{-7}$ & $5.43 \times 10^{-3}$ & 5.5 & 826  \\
C2 & 12/14/93 & 1.7 & $7.70 \times 10^{-8}$ & $2.54 \times 10^{-4}$ & 2.6 & 705  \\
C3 & 12/6/93 & 0.5 & $5.70 \times 10^{-19}$ & $1.00 \times 10^{1}$ & 49.0 & 224   \\
C3 & 12/14/93 & 0.5 & $1.47 \times 10^{-19}$ & $1.16 \times 10^{-1}$ & 13.5 & 108  \\
\end{tabular}}
\end{table}
\par One can also look at the time evolution of $B$ and $N$ for component C2.
Figures 7 and 8 plot $B$ and $N$, respectively, for the allowed
plasmoids models of C2 at all three epochs. Figures 7 and 8 show
that the plasmoid is expanding for any reasonable value of $B$ and
$N$. Furthermore, Figure 7 indicates that $B$ is decreasing as the
plasmoid propagates away from the central engine. Again, on December
14, C2 is still too small to be resolved except marginally with high
frequency with VLBA.
\section{Ramifications of a Protonic Model}
In order to discuss, the possibility of protonic plasmoids versus
positronic plasmoids, one needs to separate the energy content into
two pieces. The first is the kinetic energy of the protons,
$E(\mathrm{proton})$. The other piece is named the lepto-magnetic
energy, $E(\mathrm{lm})$, and is composed of the volume integral of
the leptonic thermal energy density, $U_{e}$, and the magnetic field
energy density, $U_{B}$. It is straightforward to compute the
lepto-magnetic energy in a spherical volume from the solutions in
Figures 5 to 8 for $B$ and $N$,
\begin{eqnarray}
 && E(\mathrm{lm}) = \int{(U_{B}+ U_{e})}\, dV = \frac{4}{3}\pi R^{3}\left[\frac{B^{2}}{8\pi}
+ \int_{1}(m_{e}c^{2})(N_{\Gamma}E^{-n + 1})\, d\,E \right]\;.
\end{eqnarray}
The kinetic energy of the protonic component is
\begin{eqnarray}
 && E(\mathrm{protonic}) = (\Gamma - 1)Mc^{2}\;,
\end{eqnarray}
where $\Gamma$ is the bulk Lorentz factor and $M$ is the mass of the
plasmoid. There is no high resolution multi-epoch radio imaging of
the ejected plasmoids for the December 1993 flare, so one cannot
estimate these kinematic values directly from observation. However,
GRS 1915+105 has shown fairly similar kinematics of the ejected
plamsoids for the various flares that have been monitored since 1994
\citep{mir94,rod95,fen99,dha00,mil05}. The kinematic fits of the
plasmoid motion for the March 1994 major flare indicate that the
motion is consistent with a bulk velocity is $v=0.92c$ and the
Lorentz factor associated with this motion is $\Gamma=2.55$
\citep{mir94,rod99}. The values are not that much different than
those inferred for other major flares \citep{fen99,rod99}. In
\citet{fen03}, it was noted that these kinematic estimates can only
be considered lower limits to $\Gamma$. However, these lower limits
are not necessarily inaccurate in every instance. With this
important disclaimer in mind, the popular March 1994 values from
\citet{mir94} for the December 1993 flare are adopted in the
following.
\par Figure 9 plots the energy
content of component C1 on December 6, 1993 for each member of the
family of models. The lepto-magnetic energy, $E(\mathrm{lm})$, is
plotted as a dashed curve for each value of minimum energy in the
adopted models $1<E_{min}<30$ as function of the modeled plasmoid
radius, $R$. The total energy, $E(\mathrm{Total})= E(\mathrm{lm}) +
E(\mathrm{protonic}) $, is plotted as a solid line. The curves are
color coded, so $E(\mathrm{lm})$ and $E(\mathrm{Total})$ are the
same color for each value of $E_{min}$. For example, the $E_{min}=1$
curves are green. First note that the minimum of $E(\mathrm{lm})$
and $E(\mathrm{Total})$ are at very different
modeled plasmoid radii. The minimum of $E(\mathrm{lm})$ occurs near
the equipartition point, $U_{B} \gtrsim U_{e}$. This is indicated by
the vertical dashed green line for $E_{min}=1$. To the left of this
line, $U_{B} < U_{e}$, and to the right of this line $U_{B} >
U_{e}$. Note that the vertical continuation of the dashed line to
its intersection with the $E(\mathrm{Total})$ curve is far to the
left of the minimum of the $E(\mathrm{Total})$ curve. Conversely, the
minimum of the $E(\mathrm{Total})$ curve occurs for models in which
$U_{B} \gg U_{e}$. One can infer that $E(\mathrm{Total})$  is
dominated by the protonic kinetic energy except for large radii
where the number density is small and the magnetic energy density is
large (i.e., the lepto-magnetic energy is much smaller than the
protonic kinetic energy unless the plasmoid is extremely
magnetically dominated). For extreme magnetic dominance,
$E(\mathrm{Total}) \approx E(\mathrm{lm})$, at large R.
\begin{figure}
\begin{center}
\includegraphics[width=125 mm, angle= 0]{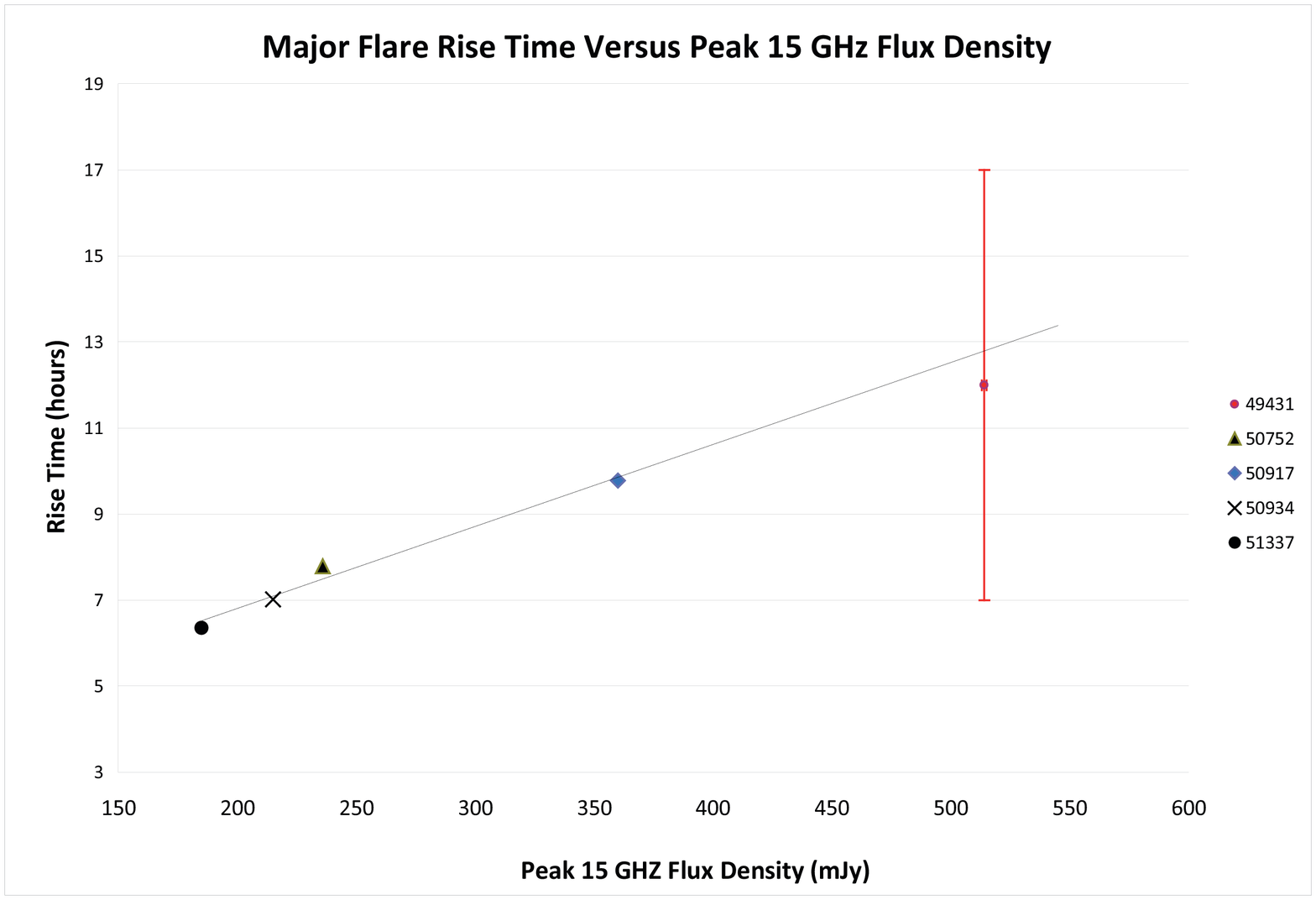}
\caption{The rise time of 4 major flares as a function of peak 15 GHz flux density
from the long term 15 GHz monitoring of \citet{rus10}.
The rise time of the flares at MJD 50917 and at MJD 50934 are derived from the dense data sampling
at 8.4 GHz and 2.3 GHz in \citet{dha00}. The crude estimate of the rise time of the March 1994 flare
from \citet{rod95} is also included. The data from the four well sampled flares is well fit by a line.}
\end{center}
\end{figure}
\par For a protonic plasma, one can compute the spatially averaged energy flux
in the plasmoid, $<Q>$, as
\begin{eqnarray}
 && <Q(\mathrm{protonic})> = (1/2R)\int_{-R}^{R}{\pi(R^{2}-r^{2})\Gamma(\Gamma-1)nvm_{p}c^{2}}\, dr \;.
\end{eqnarray}
A spatially averaged energy flux is calculated because the geometry
is spherical, not conical, by assumption. The geometrical
cross-section of the sphere that is orthogonal to the plasmoid bulk
velocity vector is zero at the poles and maximal at the equator. The
energy flux scales with this cross-sectional surface area, thus
there is major variation of the energy flux along the polar
diameter. Thus, an average over the polar diameter (projected along
the direction of the bulk velocity vector) gives a more appropriate
quantity, the spatially averaged energy flux. The $\pi(R^{2}-r^{2})$
term in equation (11) is the cross sectional area of the spheroid
that is orthogonal to the bulk velocity vector, as $r$ varies from
the north pole at $r=-R$ to the south pole at $r=R$. Similarly, for
the lepto-magnetic component of the plasma, one can compute the
spatially averaged energy flux in the plasmoid, $<Q(\mathrm{lm})>$,
as
\begin{eqnarray}
 && <Q(\mathrm{lm})> = (1/2R)\int_{-R}^{R}{\pi(R^{2}-r^{2})(\Gamma^{2})v(U_{B}+ U_{e})}\, dr \;.
\end{eqnarray}

The total energy flux, $Q(\mathrm{Total})= <Q(\mathrm{lm})> +
<Q(\mathrm{protonic})> $. This
measure of energy flux can be distinguished from the energy flux
injected into the plasmoid from the central engine. In order to
assess the origin of the plasmoids, it is much more revealing to
look at the power required to initiate the plasmoid,
$Q_{\mathrm{protonic}}(\mathrm{injected})$, rather than the
redistribution of energy flux within the plasmoid, far from the
central engine. The power injected by the central engine is defined
by
\begin{eqnarray}
 && Q_{\mathrm{protonic}}(\mathrm{injected}) = E(\mathrm{protonic})/t_{\mathrm{rise}} = (\Gamma - 1)Mc^{2}/t_{\mathrm{rise}}\;,
\end{eqnarray}
where $t_{\mathrm{rise}}$ is the amount of time for the energy to be
injected into the plasmoid, the flare rise time. Similarly, for the
lepto-magnetic energy flux that is injected
\begin{eqnarray}
 && Q_{\mathrm{lm}}(\mathrm{injected}) = E(\mathrm{lm})/t_{\mathrm{rise}}\;.
\end{eqnarray}

The radio flux density during the rise of the flare was not sampled
during the December 1993 flare. There is high time resolution radio
sampling of four major flares that was found in the archival
literature that is plotted in Figure 10.  The four flares with rise
time coverage were all monitored at 15 GHz with the Ryle telescope
in \citet{rus10}. Two of the flares were actually well sampled at 15
GHz in \citet{rus10}, the 1997 flare on MJD 50752 and the 1999 flare
on MJD 51337. The two 1998 flares in Figure 10 were not well sampled
on their rise at 15 GHz in \citet{rus10}. Fortunately, there is
excellent data sampling on the rise time at 2.3 GHz and 8.4 GHz for
the 1998 flares on MJD 50917 and MJD 50934 in \citet{dha00}. The 8.4
GHz data is used to estimate $t_{\mathrm{rise}}$ for these flares.
There is also a crude estimate on $t_{\mathrm{rise}}$ for the 1.4
GHz flux from the March 1994 flare of $12 \pm 5$ hours that was
derived in \citet{rod95} by extrapolating the motion of VLA
components backwards in time combined with sparsely sampled radio
monitoring. This point with its large error bars is also added to
the plot. The figure indicates that stronger flares tend to have
larger $t_{\mathrm{rise}}$. The strongest flare from March 1994 may
not follow the linear trend that fits the other 4 flares very well -
the error bars are too large to make this determination. The linear
fit for the other four data points is
\begin{eqnarray}
 && t(\mathrm{rise(hours)}) = 2.981 + 0.0191S_{\nu}(15 \mathrm{GHz})\;,
\end{eqnarray}
with a coefficient of determination of $R^{2} = 0.9819$. Thus, it
seems reasonable to use archival data to estimate the rise time of
the December 1993 flare. The December 1993 flares C1 and C2 have 15
GHz flux densities similar to the MJD 50752 and MJD 50934 flares, so
$t_{\mathrm{rise}} \approx 7.5 \mathrm{hrs}\approx 2 \times 10^{4}
\mathrm{sec}$ is chosen as a fiducial $t_{\mathrm{rise}}$. The
resultant $Q$ values from equations (15) and (16) can be simply
scaled with different choices of $t_{rise}$. The uncertainty here,
based on monitored flares in the rising phase seems to be a factor
of 2 at most.
\par Figure 9 contains a plot of $Q(\mathrm{Total})$ and $Q(\mathrm{lm})$ based on the values of $E$ in Figure 9 divided by
$t_{\mathrm{rise}}=2 \times 10^{4} \mathrm{sec}$. First note that
the $Q(\mathrm{Total})$ that is needed to initiate the flare in
Figure 9 is larger than the spatially averaged value of
$Q(\mathrm{Total})$ in the plasmoid when it was observed on December
6. This is expected because the plasmoid will expand spreading the
energetic particles over a larger volume, thereby lowering the
average energy flux, $<Q(\mathrm{Total})>$. Notice the color coded
vertical dashed lines in Figure 9. These lines connect the minimum
of the lepto-magnetic energy (commonly referred to as the "minimum
energy configuration" in the literature) to the corresponding
$Q(\mathrm{Total})$ for the same numerical model (determined by $R$
and $E_{min}$). For $E_{min}=1$, the total injection energy of the
"minimum energy solution" is $Q(\mathrm{Total})>
10^{43}\mathrm{ergs/sec}$ which is comparable to the strongest radio
flares in the powerful nearby quasar Mrk 231 \citet{rey09}. This
seems very unlikely. Even the "minimum energy solution" for
$E_{min}=5$ has $Q(\mathrm{Total})> 10^{41}\mathrm{ergs/sec}$,
comparable to many radio quiet quasars and their galactic hosts.
Thus, Figure 9 shows us that the combination of the minimum energy
assumption and $E_{min}<5$ is not plausible for the C1 plasmoid, if
the plasma is protonic.
\begin{figure}
\begin{center}
\includegraphics[width=125 mm, angle= 0]{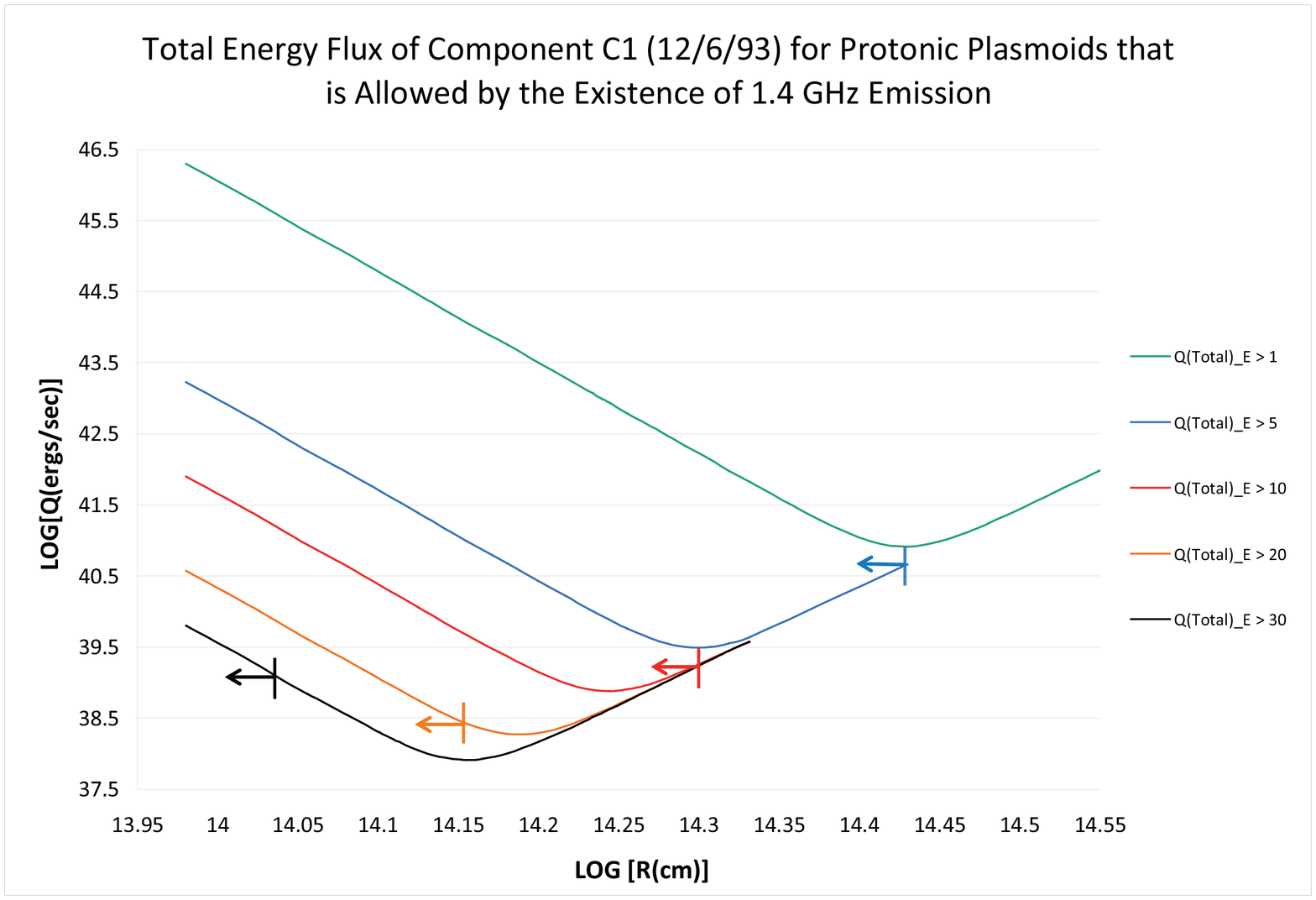}
\caption{The same plots of $Q(\mathrm{Total})$ that appear in Figure
9 with the additional constraint that the energy of the particles
responsible for the strong 1.4 GHz emission satisfy, $E(\mathrm{1.4
GHz})>E_{min}$. The horizontal partitions separate the allowed
regions of solution space from the forbidden regions of solution
space for each value of $E_{min}$. The arrows point toward the
allowed region.}
\end{center}
\end{figure}

\begin{figure}
\begin{center}
\includegraphics[width=175 mm, angle= 0]{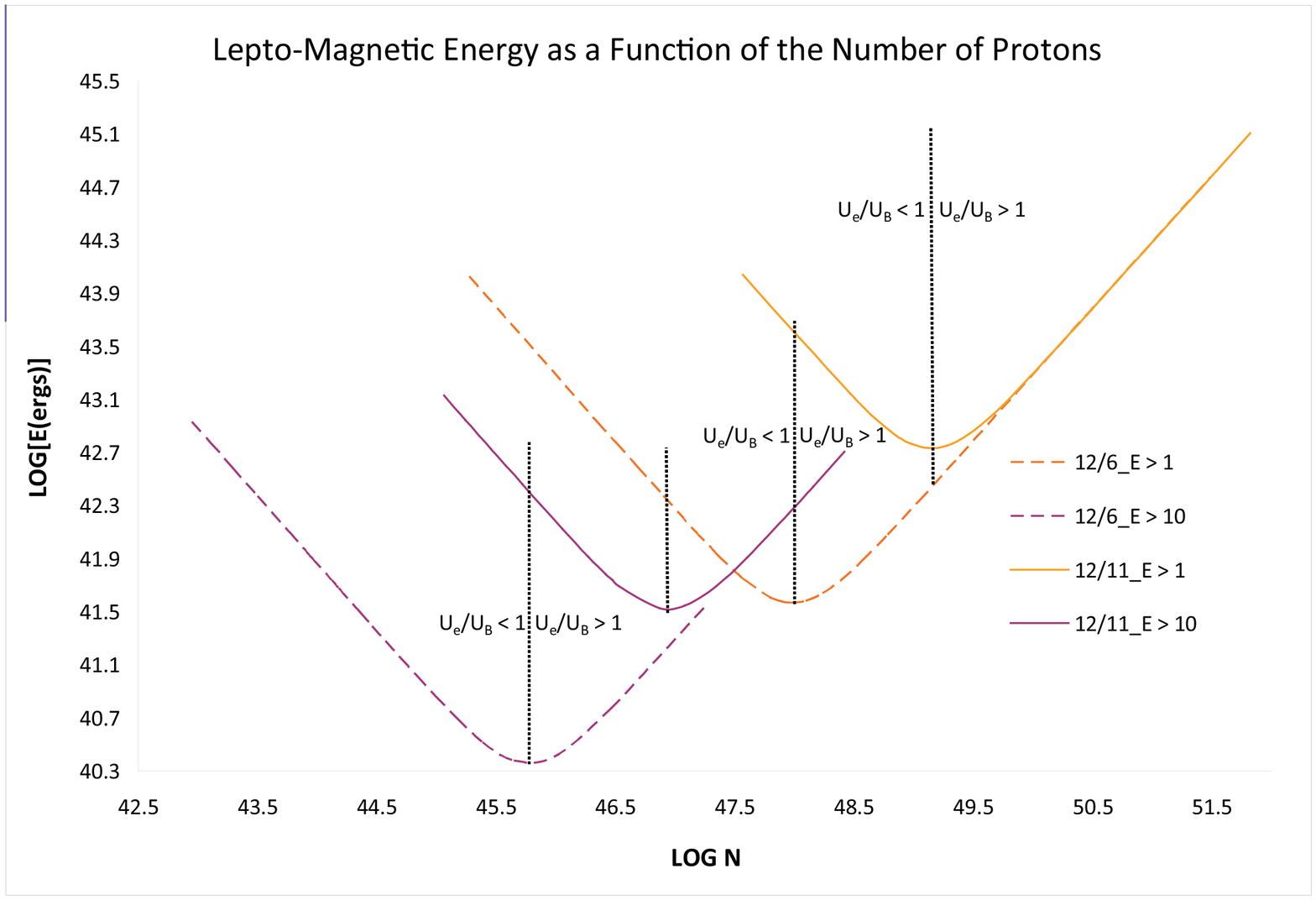}
\caption{The lepto-magnetic energy, $E(\mathrm{lm})$, in C2
on December 6 and December 11 as a function of the total number of particles, $N$, in
the model of the plasmoid for $E_{min}=1$ and $E_{min}=5$. For protonic plasmoids,
baryon number should be conserved. Thus, solutions at different epochs can be connected
by a vertical line (i.e., the dashed black vertical lines). This figure shows that
the protonic solutions require mechanical energy to be converted to magnetic energy
as the plasmoids expand.}
\end{center}
\end{figure}

\par One can use Figure 11 to restrict the solution space further.
The idea of a protonic plasma seems natural if the constituent
matter comes from the accretion flow. The large energy fluxes of the
minimum energy solutions that were noted in regards to Figure 9 are
problematic for these models. a possible resolution are solutions
that are far from minimum energy. A large magnetic field (very
non-equipartition) ameliorates the problem associated with an
enormous protonic component. However, there is a connection between
the observed emission at low frequency (1.4 GHz), the energy of the
synchrotron emitting leptons, $E(\mathrm{1.4 GHz})$, $B$ and the
constraint $E > E_{min}$. Our fits to the C1 spectra indicate that
the 1.4 GHz emission is inherently strong, but it is self-absorbed
in the magneto-plasma. Using the formula from \citet{tuc75}, one can
compute the electron energy required for a peak synchrotron flux at
an observed frequency, $\nu_{o}$, for a given magnetic field $B$,
\begin{eqnarray}
&& E_{\mathrm{peak}} \approx \left[\frac{\nu_{o}(\mathrm{peak})/\delta}{3
\times 10^{6}B}\right]^{0.5} > E_{min} \;, \quad E(1.4\,\mathrm{GHz}) \equiv
\left[\frac{1.4\, \mathrm{GHz}/\delta}{3 \times 10^{6}}\right]^{0.5}\;.
\end{eqnarray}
It is instructive to look at the restriction $E(\mathrm{1.4
GHz})>E_{min}$ applied to $Q(\mathrm{Total})$ in Figure 11. There is
no restriction within the range plotted for $E_{min}=1$. However,
there is one for the other values of $E_{min}$ that is indicated by
the color coded vertical divider and the arrow pointing towards the
allowable region of solution space. Solutions that are allowed by
equation (16) are to the left of the divider (the direction that the
arrow points), solutions that are excluded by this constraint are to
the right of the divider. The minimum allowed values of
$Q(\mathrm{Total})$ exceed the maximum sustained energy flux in the
radiation field from the accretion flow of $\sim 3 \times 10^{38}
\mathrm{ergs/sec}$ from GRS 1915+105, both contemporaneous with the
flare and historically \citep{har97}. Larger soft X-ray (possibly
thermal) luminosity has been observed, but only very briefly, not
long enough to launch the plasmoid \citep{mir98}. There is a small
region of solution space with energy fluxes comparable with the
luminosity of these high states of sustained thermal X-ray emission,
$E_{min}= 20$ restricted to a very small range of $R$ where $U_{B}
\gg U_{e}$. It seems very unlikely that disk radiation pressure can
provide the force that is necessary to initiate the ejection of the
protonic plasmoids. The implication is that magnetic forces would
likely be required to energize the protonic plasmoid and eject it at
relativistic speeds.
\par There is more information that can be gleaned from the protonic models of component C2 as they evolve in time.
This is captured in Figure 12. This figure compares the solution
space of the $E_{min}=1$ and $E_{min}=5$ solutions at two epochs,
December 6 and December 11. The solutions for other values of
$E_{min}$ were left off the chart, so as not to clutter it, but the
trends that will be described apply to all solutions. These plots
are formatted differently than what has been shown previously. This
is a plot of $E(\mathrm{lm})$ versus $N$, the total particle number
in the plasmoid (instead of $R$). As the plasmoid evolves, one
expects the baryon number to be conserved. There is no reason to
presume significant entrainment both on dynamical grounds since
Kelvin-Helmholtz instabilities are inefficient means of mixing media
for relativistic flows, \citet{bic94}, and empirically the plasmoids
in major flares do not seem to decelerate noticeably from scales of
1 mas to 100 mas \citep{dha00}. Applying baryon conservation to C2
in Figure 12, shows that as time evolves, the magnetic energy in the
system must increase. The vertical dashed lines are drawn to help
visualize this. The data is color coded. The $E_{min}=1$ data is
orange both on the December 6 curve (dashed) and the December 11
curve (solid). Similarly, the $E_{min}=5$ data is purple both on the
December 6 curve (dashed) and the December 11 curve (solid). The
vertical dashed lines connect solutions with conserved baryon number
in time. The dashed lines are chosen to intersect the four minimum
energy solutions: $E_{min}=1$, December 6; $E_{min}=1$, December 11;
$E_{min}=5$, December 6; and $E_{min}=5$, December 11. Consider the
vertical line farthest to the right that intersects the minimum
energy configuration at $E_{min}=1$, December 11. Baryon
conservation, implies that the solution is far to the right of the
minimum energy point on the plot at for $E_{min}=1$ on December 6,
where $U_{e} \gg U_{B}$. Thus, the plasmoid is evolving from a state
of $U_{e} \gg U_{B}$ to one of $U_{e} \approx U_{B}$ in 5 days.
Since radiation losses are negligible in 5 days ($\sim 10^{39}$ ergs)
compared to the mechanical energy in the plasmoid, energy
conservation implies that mechanical energy is being converted to
magnetic energy. In particular, the magnetic energy increases from
$2.69 \times 10^{40}$ ergs to $5.08 \times 10^{42}$ ergs in 5 days
in these protonic solutions.

\par One can synthesize the deductions from Figures 9, 11 and 12 as
follows:
\begin{enumerate}
\item Figure 9 shows that the kinetic luminosity of the protonic
solutions can rival quasars unless the system is far from
equipartition (specifically, it is magnetically dominated, $U_{B} \gg U_{e}$) and $E_{min} \gg 1$.
\item Figure 11 shows that the protonic solutions must be initiated
by extremely large injection energy fluxes, so they cannot
be driven by radiation pressure. Furthermore, as noted above, the plasmoid must be magnetically
dominated in order to avoid unrealistic energy fluxes. Thus, magnetic forces are
required to initiate the protonic plasmoids.
\item Figure 12 shows that mechanical energy is converted to magnetic energy in the protonic plasmoids as they evolve in time
\end{enumerate}
Points 2 and 3 are inconsistent in a straightforward interpretation.
Point 2 is that the source of energy for an ejection is magnetic
energy that is converted into mechanical form (kinetic energy and
heat) when the plasmoid is launched. Point 3 says that the opposite
must be true as the plasmoid propagates, magnetic energy is being
created from mechanical energy. Thus, there is no causal mechanism
to initiate and maintain the protonic plasmoid ejections with a
monotonic transfer of energy. One could imagine more elaborate
scenarios that are not monotonic. Ejecta begin magnetically
dominated. The energy is converted to mechanical form close to the
source then begins to convert back to magnetic form farther out.
This strong energy transfer at small distance must occur on the
orders of hours or less after the ejecta form. Such an abrupt change
in the physical state seems to be representative of a shock.
Observationally, there is presently no evidence of the entropy
generated by such shocks (radiation). Theoretically, this is not
favored either. There are two possibilities in MHD, a fast shock and
a slow shock depending on whether the flow upstream of the shock
front is above or below the Alfven speed, which is near the speed of
light in the magnetically dominated limit \citep{kap66}. In the
magnetically dominated limit, strong fast shocks are inefficient at
converting magnetic energy to mechanical energy \citep{ken84}.
Conversely, slow shocks are very efficient at converting magnetic
energy to mechanical energy. However, downstream of the shock front
the flow is necessarily slower than the sound speed
\citep{kap66,pun08}. This conflicts with all observations of major
flares. The VLBA observations of superluminal velocity for different
major flares indicate that the ejecta are relativistic even on small
scales (within hours of ejection) \cite{dha00}. Thus, a
non-monotonic scenario to make points 2 and 3 compatible is
difficult to reconcile with observation and MHD.

\par The nonexistence of protonic plasmoids in GRS 1915+105 that follows from a
synthesis of points 2 and 3 also resolves the paradox of the huge implied
energy flux described in point 1. It is interesting to note that a study of large IR flares in
\citet{fen00} determined that a protonic jet would require similarly large energy fluxes,
$Q \approx 8 \times 10^{42}$ ergs/sec. They also considered this very unlikely.

\begin{figure}
\begin{center}
\includegraphics[width=175 mm, angle= 0]{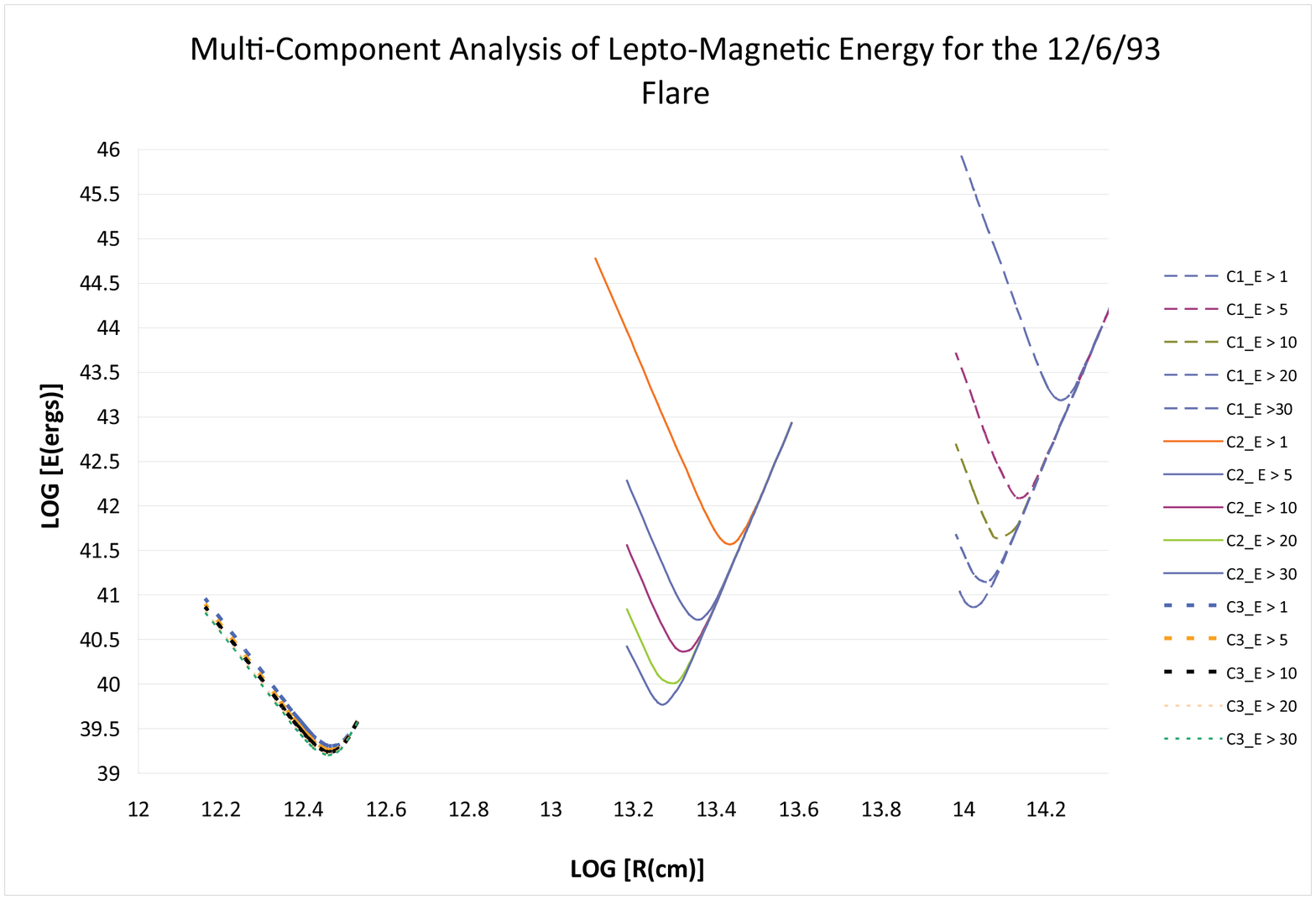}
\caption{The lepto-magnetic energy, $E(\mathrm{lm})$, in all three
components on December 6 as a function of the spheroid radius, $R$,
for the entire family of models. The lepto-magnetic energy in C3 is
very insensitive to $E_{min}$, so the curves almost overlap. It
depends more on $E_{max}$, the maximum lepton energy, because of the
flat spectrum which was not that tightly constrained by the 234 GHz
measurement (it was 12 hours out of synchronization with the the
bulk of the other data). A liberal value of $E_{max} = 10^{6}$ was
chosen. In any event, the energy content is much smaller than for C1
and C2.}
\end{center}
\end{figure}
\section{Leptonic Plasmoids}
The results of the last section indicate that physical models of the
December 1993 flare should be based on electron-positron plasmas,
not electron-proton plasmas. This section is a formal investigation
of these models. Figure 13 shows the lepto-magnetic energy stored in
each component in the models as a function of the plasmoid radius,
$R$, for the December 6 data. If one concentrates on the minimum
energy solutions then there is a curious trend that the more compact
components (younger components) have less energy. To investigate
whether this is plausible, one can also plot the lepto-magnetic
energy of the component C2 as a function of $R$ at the three
different epochs in Figure 14. Note that by December 14, the plots
of the energy stored in C2 are very similar to those of C1 on
December 6. This suggests that the energy of the flares does not
depend on the size on December 6 and this apparent trending with
size is only an artifact of the minimum energy assumption.
\begin{figure}
\begin{center}
\includegraphics[width=175 mm, angle= 0]{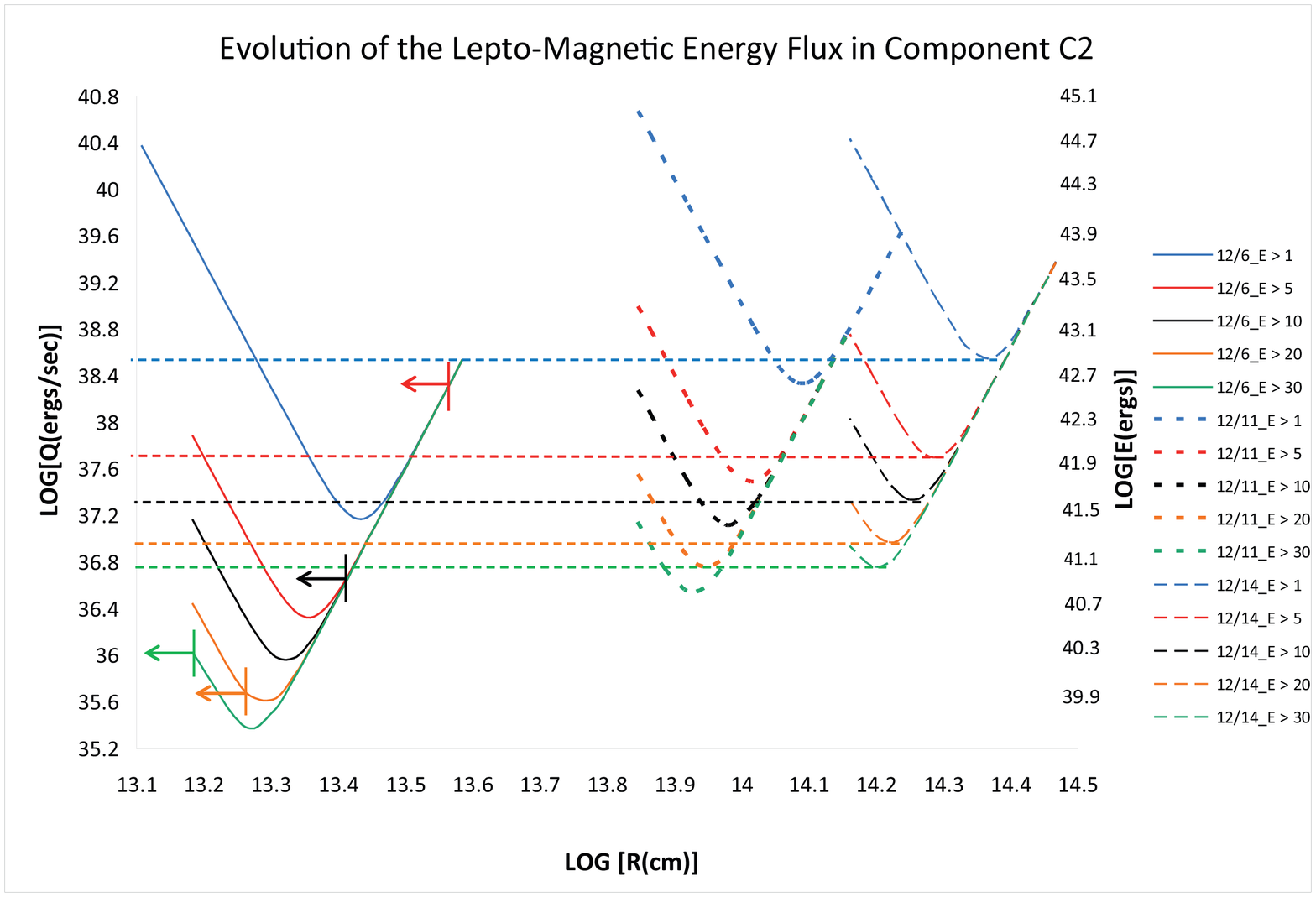}
\caption{The scale on the right hand axis describes the time
evolution in the lepto-magnetic energy, $E(\mathrm{lm})$, in C2 from
December 6 to December 14 as a function of the spheroid radius, $R$,
for the entire family of plasmoid models. Color coded horizontal
dashed lines connect color coded solutions with the same value of
$E_{min}$ and a conserved energy equal to the minimum energy on
December 14. The scale on the left hand axis quantifies the
lepto-magnetic energy flux, $Q(\mathrm{lm})$, required to launch the
leptonic plasma in C2 as deduced from the December 6 to December 14
models as a function of the spheroid radius, $R$, for the entire
family of plasmoid models. In terms of energy flux the color coded
horizontal dashed lines connect color coded solutions with the same
value of $E_{min}$ and a conserved energy flux equal to the minimum
energy flux required to launch the plasmoid based on the fits of
December 14. This allows for an evolutionary track of magnetic
energy conversion to mechanical energy as equipartition is
approached at late times. The color coded vertical partitions
separate the forbidden regions of solution space from the allowed
regions of solution space (the allowed regions are indicated by the
direction of the arrow).}
\end{center}
\end{figure}
\par Figure 14 can also be used to explore the deviations from the minimum energy assumption that were noted above.
Horizontal lines connect solutions of the same stored energy at
different times. The assumption of conserved energy seems reasonable
since the radiative losses during the 8 days of component evolution
are negligible ($\sim 10^{39}$ ergs). \footnote{There also does not
seem to be any evidence of significant adiabatic expansion energy
losses in Figure 14, so the work done by the plasmoid expanding
against an external medium seem to be negligible as well.} The plots
of the energy stored in the models of the plasmoid change mildly
from December 11 to December 14, unlike the drastic changes seen
between December 6 and December 11. This indicates that the
evolutionary rate of change of the constituent physical parameters
has slowed considerably by December 14. Furthermore, the radiative
efficiency is approaching a maximum on December 14 as noted in
regards to Figures 1 to 4. The slowing rate of change of the
physical parameters indicates that the system is close to an
equilibrium, near maximum radiative efficiency (the state of the
plasmoid is slowly approaching the minimum energy configuration).
Choosing the December 14 state as approximately the minimum energy
state seems reasonable since this is the only natural equilibrium
state of the system (of course, there could be unknown, more
complicated, dynamical principles that drive the plasmoid toward a
different equilibrium or an approach to some other asymptotic
state). This motivates the detailed study of the evolution towards
the minimum energy state that follows in this section. The results
can be applied to other asymptotic states near equipartition without
any loss of generality.
\par One can see the consequences of approximate energy
conservation in the context of the asymptotic minimum energy
assumption, by tracing the color coded horizontal dashed lines in
Figure 14. These lines connect similarly color coded solution spaces
for a particular value of $E_{min}$ at the various epochs. For
example, the blue dashed line connects the three blue $E_{min}=1$
solutions at the three different epochs maintaining a constant
energy. There are two evolutionary tracks through the solution
space. One is inertially dominated on December 6 in which mechanical
energy is converted to magnetic energy in time, until the system is
near equipartition on December 14. This was the nature of the
pathological protonic time dependent solutions that were found in
Section 4. The other time dependent solution, is one in which the
plasmoid is magnetically dominated on December 6 and magnetic energy
is converted into leptonic energy in time so that by December 14,
the system is near equipartition. One can distinguish between these
two paths through the solution space by considering two items, the
possible power sources and the exchange of energy between magnetic
and mechanical form during the course of the plasmoid evolution.
\par There are two plausible mechanisms to power the relativistic plasmoid
ejections. First, there is radiation or gas pressure from the disk
and the second is magnetic force. Figure 14 indicates that a minimum
energy configuration on December 14 evolves from a state that is far
from minimum energy on December 6. On December 6, most of the energy
is either in pair plasma or magnetic energy. If the plasmoid is
powered by radiation or gas pressure from the protonic disk, it does
not follow that it would create a highly energetic pair plasmoid.
One would expect it to be made of the same matter as its source.
However, a model of pair plasma ejected by the disk and accelerated
by the radiation field was considered in \citet{lia95}. The magnetic
field was not small initially, but was required by the equations of
motion to be significant in order to accelerate the leptons to
relativistic energies. However, note that the kinetic luminosity in
Table 3 for $E_{min}\sim 6$ is sufficiently small so as to be
consistent with the maximum sustainable radiation pressure from the
X-ray luminosity given in Section 4, $Q < 3\times 10^{38}$ ergs/sec.
Thus, the solution branch that begins with the plasmoid extremely
mechanically dominated on December 6, does not have enough magnetic
field to energize the electrons, but there is enough force from
radiation pressure to drive the asymptotic minimum energy solutions,
in principle. This solution might be viable if a suitable
acceleration mechanism in a pair plasma for the leptons were found.
Note that equation (16) implies that a weak magnetic field requires
large values of $E$ to produce flux above 10 GHz. So, the means of
accelerating the leptons is a serious technical difficulty with the
model, not just a pedantic one. The new deduction of the models
presented here is that the late time solution along the inertially
dominated branch of solution space did not evolve from one of
significant magnetic field (no lepton acceleration mechanism), so
the \citet{lia95} solution does not describe the December 1993
flare.
\par If the relativistic plasmoids are magnetically initiated then it follows
that the plasmoids should be magnetically dominated at early times.
A scenario in which a plasmoid starts out magnetic then converts to
pair plasma then back to equipartition seems very contrived and has
no known physical basis. The only causal scenario for plasmoid
evolution is one in which an energetic plasmoid is created by
magnetic forces. The magnetically dominated object dissipates its
energy into violent shocks, nonlinear MHD waves and high power
reconnection events, thereby converting magnetic energy into a
relativistically hot pair plasma.
\begin{figure}
\begin{center}
\includegraphics[width=175 mm, angle= 0]{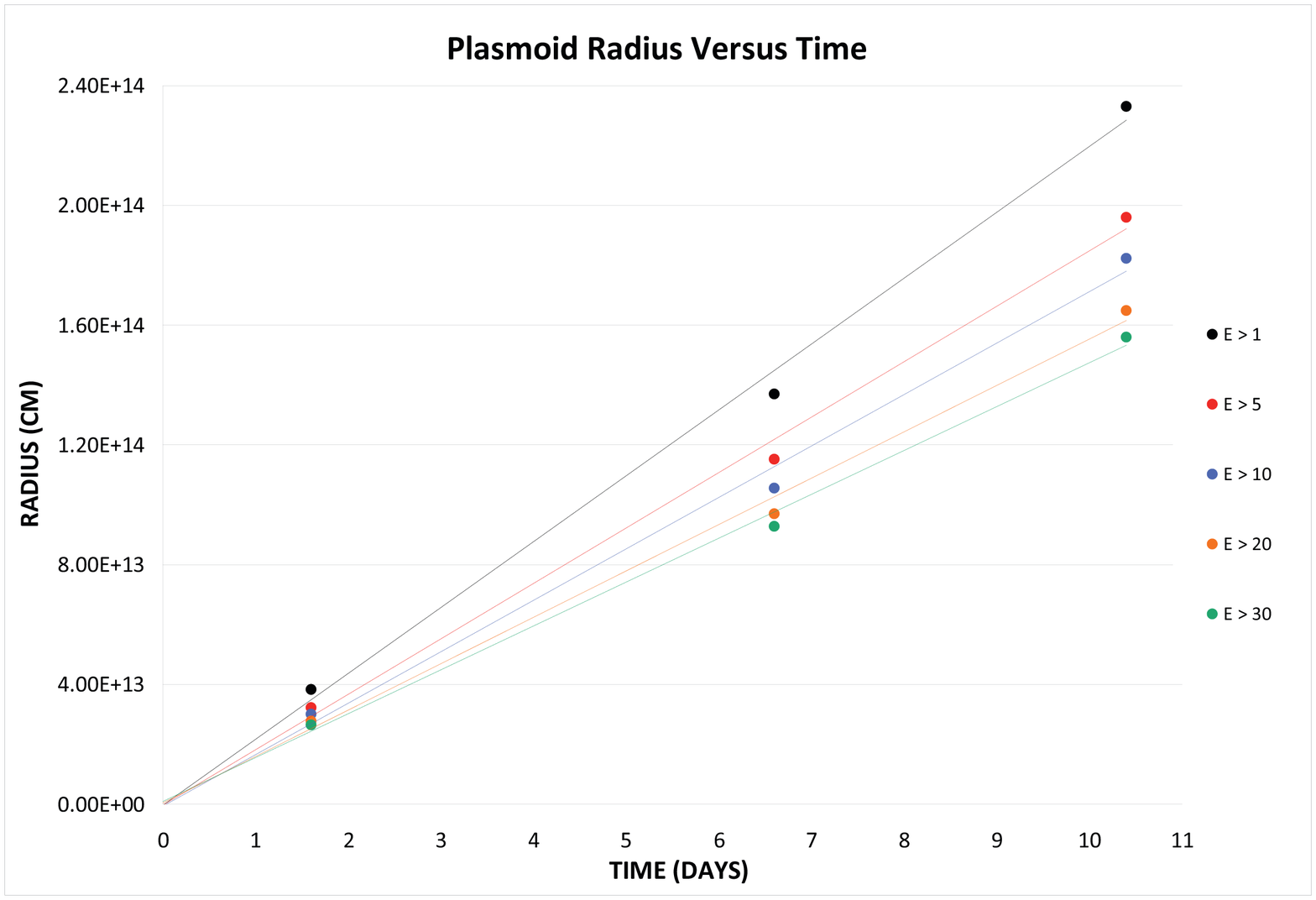}
\caption{The radius of the family of models for C2 as a function of
time. The zero point offset was adjusted so that the extrapolation
of the fits to the data reach zero radius at zero time.}
\end{center}
\end{figure}
\par Thusly motivated, one can explore the magnetically dominated time evolution track
through solution space in detail. Figure 15 is a plot of the radius
of C2 versus time for the magnetic solutions that conserve energy
and approach the minimum energy solution on December 14 as indicated
in Figure 14. The solutions follow a track of uniform expansion very
closely. The time offset was adjusted in such a way that all the
solutions reach approximately zero radius at time zero. All values
of $E_{min}$ yield virtually the same offset, indicating that the
plasmoid was 1.6 days old on December 6. For $E_{min}$=1, 5, 10, 20,
30, the least squares fit to the expansion rate, $dR/dt$, is
0.0085c, 0.0082c, 0.0078c, 0.0069c and 0.0064c, respectively.
\par One can perform an analysis similar to that which was done for Figure 11 for the protonic plasmoids
using equation (16) to restrict the allowable range of $E_{min}$.
Figure 14 is used to illustrate this and also plot the power,
$Q(\mathrm{lm})\equiv Q_{\mathrm{lm}}(\mathrm{injected})$, required
by the central engine to energize and eject the lepontic plasmoids
at relativistic velocities, where the conversion from $E$ is given
by equation (14). The plots of $Q(\mathrm{lm})$ are color coded so
that each model is the same color at all three epochs. Recall that
the restriction discussed in Section 4 arises because large values
of $B$ require small values of $E(\mathrm{1.4 GHz})$ to create a
synchrotron peak near 1.4 GHz. The spectral fits indicate that the
1.4 GHz flux is large and only appears suppressed because of
synchrotron-self absorption (i,e, the background power law spectrum
should extend lower than 1.4 GHz). The restriction equates to
$E(\mathrm{1.4 GHz})>E_{min}$. As for Figure 14, the color coded
short vertical lines and arrows indicate the radius of the model
such that $B$ is sufficiently large that equation (16) implies,
$E(\mathrm{1.4 GHz})=E_{min}$. The arrow points to the left, the
region of smaller $B$, where $E(\mathrm{1.4 GHz})> E_{min}$ and the
solutions are allowed. To the right of the vertical partition, $B$
is larger, therefore $E(\mathrm{1.4 GHz})< E_{min}$ and the
solutions are forbidden. These partitions were added to the December
6, 1993 plots. In the the evolutionary track in which magnetic
energy is converted to mechanical energy, the magnetic field is
largest at this epoch and therefore more likely to provide the
tightest constraint on $E_{min}$. In summary, the solution space is
restricted by three constraints,
\begin{enumerate}
\item The solution is magnetically dominated on December 6
\item $E(\mathrm{1.4 GHz})> E_{min}$
\item Energy conservation with the December 14 solution, $E(\mathrm{December\, 14})=E(\mathrm{December\, 11})=E(\mathrm{December\, 6})$
\end{enumerate}

Figure 14 shows that the $E_{min}=20$ and $E_{min}=30$ allowable
solutions on December 6 have lower energy than the minimum energy
solution on December 14 unless they are dominated by pair plasma
inertia (far to the left of the energy minimum on December 6). This
violates condition 1 above, so these solutions do not appear to be
physical. The $E_{min}=10$ allowable solutions on December 6 can be
magnetically dominated as indicated by the location of the black
arrow (to the right of the minimum energy). However, the black arrow
is below the black horizontal dashed line indicating that the energy
flux of these solutions is below that of the minimum energy solution
on December 14, so they would not satisfy energy conservation
(violates condition 3 above). Thus, there are no physically allowed
solutions with $E_{min}=10$. The red vertical arrow for the
$E_{min}= 5$ allowable solutions lies above the horizontal red
dashed line, so there is an allowed solution that is magnetically
dominated on December 6 and the energy is conserved until December
14. A careful numerical calculation shows that the largest value of
$E_{min}$ that is allowed by these arguments is $E_{min}=6$. Note
that departures from minimum energy on December 14, will only drop
the maximum allowed value of $E_{min}$ below 6. So, this argument is
more general than one associated with just an asymptotic minimum
energy state. The maximum allowed value of $E_{min}$ is 6 for any of
the possible magnetically dominated tracks through the solution
space in Figure 14 (i.e., other horizontal tracks that are not
indicated by the dashed lines).

\par Table 3 displays the parameters of solutions for C2 that approach minimum
energy on December 14 and for C1 on December 6. The table is a
useful device for understanding the time evolution of the plasmoid
C2 and its relationship to C1 at late times. Columns 1 and 2 in
Table 3 identify the flare. The next two columns are $E_{min}$ and
the designation of the solution as minimum energy or not,
respectively. Columns 5 and 6 are the radius of the plasmoid model
and the magnetic field strength, respectively. Column 7 is the ratio
of magnetic energy density (trivially computable from column 6) to
leptonic energy (i.e., the degree of magnetic dominance). Column 9
is the total number of particles in the plasmoid. The last column is
the power required to energize and eject the plasmoid at
relativistic speed from the central engine as defined in equation
(14).

\begin{table}
\caption{Parameters of Models that Approach Minimum Energy at Late Time Models}
{\footnotesize\begin{tabular}{cccccccccc} \tableline\rule{0mm}{3mm}
Component &  Date & $E_{min}$ & Minimum & $R$ & $B$ & $U_{B}/U{e}$ & $T_{b}$ & $N$ & $Q$\\
          &       &           &  Energy & cm & Gauss &  & $10^{11}\,^{\circ}$K &  & ergs/sec\\
\tableline \rule{0mm}{3mm}
C1   &  12/6/93  & 1 & Yes & $1.72 \times 10^{14}$ & 4.2 & 1.27 & 1.91 & $4.93 \times 10^{49}$ & $6.14\times 10^{38}$\\
C2   &  12/6/93  & 1 & No & $3.83 \times 10^{13}$ & 42.5 & $2.24 \times 10^{3}$ & 0.62 & $2.21 \times 10^{46}$ & $3.56\times 10^{38}$\\
C2   &  12/11/93  & 1 & No & $1.37 \times 10^{14}$ & 5.9 & 9.94 & 1.61 & $4.39 \times 10^{48}$ & $3.56\times 10^{38}$\\
C2   &  12/14/93  & 1 & Yes & $2.30 \times 10^{14}$ & 2.0 & 1.18 & 2.89 & $2.10 \times 10^{49}$ & $3.56\times 10^{38}$\\
C1   &  12/6/93  & 6 & Yes & $1.33 \times 10^{14}$ & 1.5 & 1.23 & 3.19 & $5.01 \times 10^{47}$ & $3.68\times 10^{37}$\\
C2   &  12/6/93  & 6 & No & $3.14 \times 10^{13}$ & 18.0 & $1.89 \times 10^{3}$ & 0.92 & $4.32 \times 10^{44}$ & $4.05\times 10^{37}$\\
C2   &  12/11/93  & 6 & No & $1.10 \times 10^{14}$ & 2.5 & 6.43 & 2.50 & $1.05 \times 10^{47}$ & $4.05\times 10^{37}$\\
C2   &  12/14/93  & 6 & Yes & $1.91 \times 10^{14}$ & 0.9 & 1.11 & 4.31 & $4.10 \times 10^{47}$ & $4.05\times 10^{37}$\\
\end{tabular}}
\end{table}
The component sizes in Table 3 are much smaller than normally
assumed in the literature and compactness might be an issue. There
might be significant inverse Compton emission that could affect the
basic assumptions of the model or further constrain the models. A
simple measure of the significance of strong inverse Compton cooling
is the brightness temperature, $T_{b}$. These are evaluated for the
various models in column (8) of Table 3 by the following method.
Physically, it is $T_{b}$ evaluated in the frame of reference (the
intrinsic brightness temperature) of the plasmoid, $(T_{b})_{p}$,
that is relevant for assessing a possible "inverse Compton
catastrophe" \citep{kel69,mar79}. We want to express this in terms
of observable quantities at earth designated by the subscript "o."
First of all, the brightness temperature is the equivalent blackbody
temperature of the radiation assuming one is in the Planck regime,
$h \nu \ll k_{b}T$. Consider a source in which the monochromatic
intensity is $I(\nu)_{p}$ from a spheroid of radius R. The
brightness temperature associated with the spectral luminosity is
\begin{eqnarray}
&& (T_{b})_{p}=\frac{S(\nu)_{o}\delta^{-1}\mathrm{c}^{2}}{2\Omega_{o}
k_{b}\nu_{o}^{2}}\;,
\end{eqnarray}
where $\Omega_{o}$ and $(S_{\nu})_{o}$, are the solid angle
subtended by the source and the flux density, respectively. It was noted in
\citet{rea94} that blazar jets tend to have $(T_{b})_{p}\approx
10^{11}\,^{\circ}$K. Thus, Table 3 indicates that the values of
$(T_{b})_{p}$ are basically consistent with those expected in
powerful astrophysical relativistic jets per the analysis of
\citet{rea94}. This value is below $(T_{b})_{p}\approx
10^{12}\,^{\circ}$K where the inverse Compton catastrophe occurs as
a consequence of the inverse Compton radiation lifetime becoming
much shorter than the synchrotron radiation lifetime \citep{kel69}.
The ratio of inverse Compton luminosity, $L_{ic}$, to synchrotron
Luminosity, $L_{synch}$ was calculated in \citet{rea94} as
\begin{eqnarray}
&& \frac{L_{ic}}{L_{synch}} =
\left(\frac{T_{b}G(\alpha)^{1/5}}{10^{12.22}}\right)^5 \left[1 +
\left(\frac{T_{b}G(\alpha)^{1/5}}{10^{12.22}}\right)^5 \right]\;,\\
&& G(\alpha) =\\ \nonumber
&&\left[\frac{v_{\mathrm{op}}/\delta}{3.5\, \mathrm{GHz}}+
\frac{v_{\mathrm{op}}/\delta(1 - \alpha)}{1\, \mathrm{GHz}}\times
\left[\left(\frac{v_{\mathrm{high}}}{v_{\mathrm{op}}}\right)^{(1
-\alpha)}-1 \right] \right] f_{3}(\alpha)^{-4}\;,
\end{eqnarray}
where $v_{\mathrm{op}}$ is the peak in the observed SSA spectrum,
$v_{\mathrm{high}}$ is the highest frequency in the power law in the
observers frame. This expression also requires the coefficient,
$f_{3}(\alpha)$, that is given by equation (15c) of \citet{sch68}.
In Table 4, the importance of the inverse Compton emission to the
individual models is considered.

\begin{table}
\caption{Inverse Compton Losses in the Component Models}
{\footnotesize\begin{tabular}{ccccccccc} \tableline\rule{0mm}{3mm}
Component &  Date & $E_{min}$ & Minimum & $v_{\mathrm{op}}$ & $v_{\mathrm{high}}$ & $\alpha$ & $T_{b}$ & $L_{ic}/L_{synch} $ \\
          &       &           &  Energy &  GHz               &   GHz              &              & $10^{11}\,^{\circ}$K &  \\
\tableline \rule{0mm}{3mm}
C1   &  12/6/93  & 1 & Yes & 2.9 & 50 & 2.2 & 1.91 & $3.23 \times 10^{-1}$ \\
C2   &  12/6/93  & 1 & No & 9.5 & 50 & 1.7 & 0.62 & $1.40 \times 10^{-5}$ \\
C2   &  12/11/93  & 1 & No & 5.5 & 50 & 1.7 & 1.61 & $1.06 \times 10^{-3}$ \\
C2   &  12/14/93  & 1 & Yes & 2.6 & 50 & 1.7 & 2.89 & $1.01 \times 10^{-2}$ \\
C1   &  12/6/93  & 6 & Yes & 2.9 & 50 & 2.2 & 3.19 & $1.45 \times 10^{1}$ \\
C2   &  12/6/93  & 6 & No & 9.5 & 50 & 1.7 & 0.92 & $1.01 \times 10^{-4}$ \\
C2   &  12/11/93  & 6 & No & 5.5 & 50 & 1.7 & 2.50 & $9.48 \times 10^{-3}$ \\
C2   &  12/14/93  & 6 & Yes & 2.6 & 50 & 2.2 & 4.31 & $8.03 \times 10^{-2}$ \\
\end{tabular}}
\end{table}

The values of $L_{ic}/L_{synch}$ in Table 4 represent a possible
constraint on the solution spaces presented here. The ratios are not
that sensitive to $v_{\mathrm{high}}$ which is not determined by the
available observations, so an arbitrary value of 50 GHz is chosen.
It was noted in \citet{rea94} that AGN tend to cluster at values of
$T_{b}$ for which $L_{ic}$ starts to compete with $L_{synch}$,
$(T_{b})_{p}\gtrsim 10^{11}\,^{\circ}$K. This concept is consistent
with the values in Table 4. The one exception is C1, with the value
of $E_{min}=6$, that appears to produce too much inverse Compton
emission to be consistent with this concept. Thus, the results in
Table 4 seem to favor the lower $E_{min}$ range in the C1 solution
space. Otherwise, it is concluded that the inverse Compton losses in
Table 4 seem reasonable for a relativistic plasmoid.

\par According to Table 3, the maximum injected energy flux of component C1 is larger than
that of component C2. Also, the larger minimum bound on the
injection energy flux in Table 3 is for C2. Thus, the maximum
injected power during the December 1993 flare is bounded to the
range $4.05 \times 10^{37}\mathrm{erg/sec}<Q(\mathrm{December, \,
1993})_{\mathrm{max}}< 6.14 \times 10^{38}\mathrm{erg/sec}$. Assume
that there are oppositely directed plasmoids that are ejected
simultaneously as commonly occurs in other observed flares. These
components typically comprise only $\sim$ 10\% of the total flux
density due to the more intense redshifting when the ejections are
directed away from the observer \citep{rod99,fen99,mil05}. It is
noted that the estimates of $Q$ for the components C1 and C2
presented here are slightly high since this extra 10\% of the flux
appeared in their component flux densities. The existence of
counter-ejecta would double the estimate above for the maximum total
injected power from the central engine during the December 1993
flare, $9.1 \times 10^{37}\mathrm{erg/sec}<Q(\mathrm{December\,
1993})_{\mathrm{max}}< 1.22 \times 10^{39}\mathrm{erg/sec}$. The
central black hole in GRS 1915+105 has a mass estimated at 14
$M_{\odot}$ \citep{gre01}. Thus, the maximum impulsive power
required to initiate the plasmoids is
 $0.05 L_{\mathrm{Edd}}< Q_{\mathrm{max}} < 0.68 L_{\mathrm{Edd}}$ in terms of the Eddington luminosity,
$L_{\mathrm{Edd}}$.

\section{Discussion} In this paper, a major flare from December 1993 from the micro-quasar
GRS 1915+105 was analyzed in order to address four major unknowns
from historical estimates of flare energy flux. These major
uncertainties that were discussed in the Introduction are listed
again here with the corresponding resolution implied by the detailed
modeling described in the previous sections.
\begin{enumerate}
\item \textbf{Is the plasma protonic?} The answer seems to be no, it is electron-positron plasma. In
section 4, it was shown that the energy fluxes required by the
electron-proton plasma assumption exceed the radiated energy flux
from the accretion flow (see Figure 11), implying that the plasmoids
are driven by magnetic forces. Yet, the time evolution of C2 (see
Figure 12) implies that mechanical energy is converted to magnetic
energy over time. The time evolution history of this scenario is
internally inconsistent. Thus, there is no causal time evolution if
the protonic assumption is invoked.
\item \textbf{What is the minimum electron energy, $E_{min}$?} In section 5,
it was shown that constraints imposed by the large magnetic fields
in the compact plasmoids imply (through equation (16)) that small
values of $E$ can produce the observed 1.4 GHz flux densities. Since
one must have $E > E_{min}$, large values of $E_{min}$ are
forbidden. For the models of component, C2, $1 < E_{min} < 6$.

\item \textbf{There is uncertainty in the size of the region that produces the bulk of the radio
emission.} In section 5, the models are used to show that the reason
that interferometry does not resolve the preponderance of major
flare flux in GRS 1915+105 is that the plasmoids responsible for
this emission are $\sim$ 10 AU. The small plasmoid sizes are
physically self-consistent as evidenced by the brightness
temperatures of the plasmoids in Table 3. The volume used in
previous estimates of the energy flux are typically two to three
orders of magnitude too large.
\item \textbf{Is the minimum energy or the equipartition assumption justified?} Definitely not at early times,
but possibly at late times. Figure 14 shows that there is no minimum
energy evolutionary track through the solution space for C2 from
December 6, 1993 to December 14, 1993.
\end{enumerate}
The study presented here indicates that the major flares in GRS
1915+105 are discrete ejections of highly magnetized
electron-positron plasma. The flares are associated with a strong
long term magnetic transient that resides at the heart of the
central engine that lasts $\sim 4.6\times 10^{7}$ black hole spin
periods (assuming a rapidly spinning black hole). As the flares
propagate, they expand with a radial velocity $\approx 0.008c$
converting magnetic energy into a relativistically hot pair plasma.
The models were able to bound the power required to energize and
launch the most energetic flares, $8.10 \times
10^{37}\mathrm{erg/sec}<Q(\mathrm{December\, 1993})< 1.22 \times
10^{39}\mathrm{erg/sec}$. This result agrees very closely with the
minimum energy estimate of \citet{gli99} for the March 1994 flare,
if the radius of the emitting volume is corrected. They assumed a
size of $R = 7 \times 10^{15}$ cm, which is a value much less than
the interferometer beamwidth. This sub-beamwidth size was derived
from a Gaussian fit algorithm to the size of the source (the
unresolved core plus secondary) that was determined in
\citet{rod99}. However, the model fits presented here indicate that
the true size of the preponderance of the plasmoid emission should
be much less than this. It is argued here that $R$ is likely to have
been overestimated by a factor of 10 to 50 (based on the age of the
flare when it was observed, $\sim$ 5 days, and the results
summarized in Table 3). The VLA at 8.3 GHz cannot resolve a region
of high emissivity $\sim 3 \times 10^{14}$ cm buried within a
larger, inhomogeneous, low luminosity region of $\sim 7 \times
10^{15}$ cm. From equation (5) of \citet{gli99}, $Q \sim R^{0.5}$,
thus the adjusted $Q$ values for an electron-positron plasma in
\citet{gli99} are $3.0 \times
10^{38}\mathrm{erg/sec}<Q(\mathrm{March\, 1994})< 9.2 \times
10^{38}\mathrm{erg/sec}$. They also considered magnetic dominance as
a possible state of the ejected plasma, but did not pursue it in
much detail. A second consistency check is provided by the strong
flare observed with MERLIN on MJD 50752.1 in \citet{fen99}. They
assumed a flare rise time of 12 hours and a plasmoid size of
$10^{15}$ cm which is less than 0.14 of the MERLIN FWHM beamwidth.
The plasmoid was unresolved in the image plane. Using Figure 10 to
adjust the flare rise time to 8 hours and Table 3 to adjust the
plasmoid radius to $\approx 2 \times 10^{14}$ cm, equation (6) of
\citet{fen99} and Table 3 of \citet{fen99} yield an "adjusted"
minimum injection energy flux of $Q \approx 3.8 \times
10^{38}\mathrm{erg/sec}$ for an electron-positron plasmoid, similar
to the values presented here.

\par This study illustrates the investigative power provided by densely sampled broadband radio data
in the study of major flares in GRS 1915+105. The power of this
method is a strong argument to initiate a study program that
involves broadband quasi-simultaneous spectra (simultaneous to
within a few hours and extending above 100 GHz) coordinated with
high frequency VLBA observations (15 GHz and 22 GHz).

\end{document}